\documentclass[reprint, superscriptaddress, amsmath, amssymb, twocolumn, pra, nobibnotes]{revtex4-2}

\usepackage{amsmath}
\usepackage{mathtools}
\usepackage{amssymb}
\usepackage{graphicx}
\usepackage{bm}
\usepackage[colorlinks,linkcolor=blue,citecolor=blue,urlcolor=blue,breaklinks=true]{hyperref}
\usepackage{booktabs}
\usepackage{multirow}
\usepackage[dvipsnames]{xcolor}
\usepackage{ragged2e}
\usepackage{orcidlink}
\usepackage[T1]{fontenc}

\graphicspath{{Figs/}}

\newcommand{\kb}{k_{\rm B}}
\newcommand{\Tr}{\operatorname{Tr}}

\newcommand{\figpanel}[2]{\hyperref[#1]{\ref*{#1}(#2)}}
\newcommand{\ket}[1]{\left|#1\right\rangle}

\begin{document}

\title{Finite Quantum Engineering Thermodynamics}

\author{Borhan Ahmadi\orcidlink{0000-0002-2787-9321}}
\email{borhan.ahmadi@ug.edu.pl}
\affiliation{International Centre for Theory of Quantum Technologies,
University of Gda\'nsk, ul.~prof. Marii Janion 4,
80-309 Gda\'nsk, Poland}

\date{\today}

\begin{abstract}
A useful thermodynamic model predicts what a device will do without tracking its full microscopic state. For a finite quantum working medium, the difficult question is which quantities should actually be measured. We develop a task-based method that keeps exact energy accounting separate from calibrated macroscopic prediction. Operator geometry identifies what a measured record determines and bounds what remains uncertain. We then compare direct measurements of dynamical observables with earlier measurements of the same coarse record, selecting the representation together with its measurement cost. A Bose--Hubbard example yields a compact response table tested over a continuous operating region and on unseen microscopic trajectories. A driven Ising chain shows why history is not always the cheaper choice. The result is a way to turn detailed quantum dynamics into a small, reusable, and explicitly qualified thermodynamic description.
\end{abstract}

\maketitle

\section{Introduction}
\label{sec:introduction}

Thermodynamics is useful because it tells us how a system will behave without following every microscopic degree of freedom. At equilibrium, a small set of independent properties determines the remaining equilibrium properties~\cite{Fermi1956,Cengel2024,KlineKoenig1957}. Away from equilibrium, exact balance laws still account for energy and other conserved quantities, but they do not determine the response of a particular material. That response must be supplied by constitutive relations~\cite{Noll1958,ColemanNoll1963,Muller1986}. A practical description is therefore intentionally incomplete: it keeps the information needed for its purpose and uses constitutive knowledge to account for the rest.

Classical nonequilibrium thermodynamics developed several ways to make this separation work. Onsager related near-equilibrium fluxes to their driving forces~\cite{Onsager1931a,Onsager1931b,deGrootMazur1962}. For more general processes, the present macroscopic variables may not determine the future response. One can then add internal variables or fluxes to the instantaneous state, or allow the constitutive relation to depend on earlier history~\cite{GreenRivlin1957,Coleman1964Memory,ColemanGurtin1967,Day1972,Muschik1990,Muschik2021}. Classical rational thermodynamics also asks whether such relations obey physical restrictions, including those imposed by the entropy inequality~\cite{ColemanNoll1963,ColemanGurtin1967}. Predictive accuracy alone is not the whole of constitutive thermodynamics.

This classical work also sharpened the meaning of a state. In theories of materials with memory, different histories can be treated as the same state when they give the same response to the relevant future processes~\cite{Noll1972,Willems1972,FabrizioMorro2005,DeseriFabrizioGolden2006}. A related lesson comes from statistical mechanics: when microscopic variables are left unresolved, their influence generally returns through residual forces or memory~\cite{Zwanzig1960,Zwanzig1961,Mori1965,Grabert1982}. Optimal-prediction methods and studies of partially observed systems have developed this idea further, showing how unresolved dynamics can be represented statistically or through past measurements~\cite{Chorin2000,Chorin2002,Rupe2022}. Predictive states and the choice of informative measurements also have established foundations in computational mechanics and control~\cite{ShaliziCrutchfield2001,ZhangAyoubSundaram2017}. We build on these insights; neither memory nor the search for a smaller predictive state is, by itself, new.

For a finite quantum working medium, however, the practical choice of state has an extra cost. The density operator gives the full statistical description, but routine tomography can require far more measurements than the task itself. Different observables may also require incompatible measurement settings. We must therefore decide not only which information helps prediction, but whether obtaining that information is worth the experimental effort. Throughout, the microscopic state, the chosen system boundary, and the measured thermodynamic record remain distinct. In particular, heat and work describe physical energy transfer, while an information-restricted entropy change is not automatically physical entropy production~\cite{Fermi1956,Cengel2024,ahmadi2020NoEntropy}.

Our earlier restricted-information studies addressed two parts of this problem. Reference~\cite{Ahmadi2026Endogenous} began with a fixed coarse record and showed how unresolved microscopic currents can govern its motion. Reference~\cite{Ahmadi2026FiniteParticle} then combined a prescribed spatial record with mean energy to support thermodynamic accounting and a complete cycle without microscopic state reconstruction. In both cases, the accessible information was chosen at the outset.

That leaves the main engineering question unanswered. Given a working medium, a set of controls, and a required prediction, \emph{which quantities should actually be measured}? The present record may be too small to predict its own future. Additional current-like observables might resolve that problem, but they can be costly to measure. A few earlier measurements may be easier, but closely spaced history can magnify statistical noise. Nor is a good fit at one operating point enough: the representation must remain reliable over its declared range of use. The challenge is to choose the accessible state, its constitutive response, and its measurement budget together.

Here we make that choice quantitative for finite quantum working media. Exact operator relations tell us what a proposed record already fixes and bound the ambiguity left by unmeasured information. Microscopic dynamics is then used before operation to calibrate the remaining response and assess its range of validity. The selected measured state is the one that meets the declared prediction task at an acceptable cost within the candidate families. A Bose--Hubbard working medium demonstrates a fixed, compact constitutive data sheet and an independent numerical holdout. A driven Ising chain shows that the preferred choice can change with the forecast horizon. The physical distinction is not simply between memory and no memory: a useful past record can carry directions outside a \emph{truncated} local derivative space, or it can become a noisy way of measuring those same local directions.

Our aim is thus a task-certified constitutive prediction framework, not a general replacement for the entropy-based admissibility conditions of classical dissipative thermodynamics. Detailed quantum dynamics supplies the constitutive information offline; routine operation needs only the small measured state and the already tested response. This is how the classical separation between state, balance law, and constitutive knowledge becomes an explicit quantum measurement and prediction procedure.

\section{Thermodynamic prediction from restricted quantum information}
\label{sec:framework}
%
\begin{figure}[t]
\centering
\includegraphics[width=\linewidth]{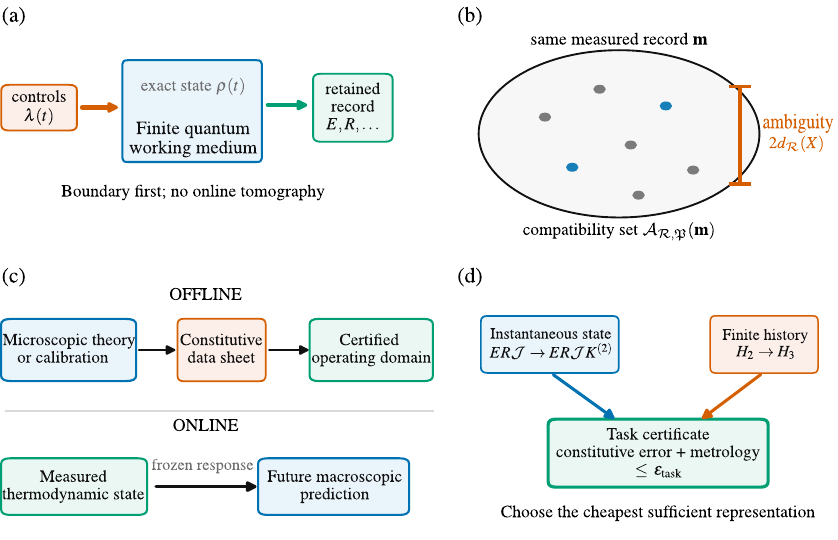}
\caption{\textbf{Finite quantum engineering thermodynamics.}
\textbf{(a)} The working medium, its boundary, and the controls are fixed before selecting what to measure. The microscopic state is \(\rho(t)\), but routine operation retains only a small record.
\textbf{(b)} Distinct microscopic states can give the same record. Their possible differences in a requested output define the unresolved prediction ambiguity.
\textbf{(c)} Detailed dynamics is used offline to fit and test a constitutive response; online prediction then combines this fixed response with the measured record, without tomography.
\textbf{(d)} Missing motion can be probed through present currents and derivatives or through earlier records. The choice balances unresolved-state error against the statistical cost of measuring the added information.}
\label{fig:framework}
\end{figure}

The earlier restricted-record results explain why a coarse state can be dynamically incomplete. We now turn that observation into an engineering construction. The goal is to separate what must be measured online from what can be learned once, offline, from the microscopic model, and then to state clearly when the resulting prediction is reliable. The connection to our earlier results is recalled briefly (see Note~\ref{supp:note-lineage} of the SM \cite{SuppMat} for the lineage of the restricted-record framework).

This division of labor is depicted in Fig.~\figpanel{fig:framework}{a}--\figpanel{fig:framework}{c}. The physical working medium and its controls are fixed first; the measured record then identifies a set of compatible microscopic states; microscopic quantum dynamics is finally used offline to construct the response needed for online prediction. The fourth panel, discussed below, shows how the representation itself is selected when the present record does not contain enough dynamical information.

\subsection{Physical boundary and thermodynamic record}
\label{subsec:record}

We first specify the physical system. The working medium, its boundary, and the externally controlled parameters are chosen before heat, work, or entropy transfer are discussed. This is important because heat and work are modes of energy transfer across the chosen boundary, not properties stored in the
system. During a thermally isolated Hamiltonian stroke, \(\delta Q=0\), while a controlled change \(d\lambda_a\) transfers work
\(\delta W_{\rm on} = \langle\partial H/\partial\lambda_a\rangle d\lambda_a\) to the system \cite{Fermi1956,Cengel2024}. The reduced description introduced below does not change these physical definitions. We calibrate observable predictions for isolated strokes; we do not introduce a phenomenological dissipative law or claim that a prediction-error certificate replaces the entropy restrictions used in classical constitutive thermodynamics~\cite{ColemanNoll1963,ColemanGurtin1967}.

We next decide what information is retained experimentally. Let
\(\mathcal R=\{A_1,\ldots,A_m\}\) denote a set of measurable Hermitian observables. Their expectation values
\(m_\mu=\Tr(\rho A_\mu)\) form the measured record
\(\mathbf m=(m_1,\ldots,m_m)\). The density operator \(\rho\) remains the exact microscopic state, but it is not reconstructed during online operation. The thermodynamic description uses only the much smaller measured record, together with the externally known control parameters.

A measured record does not normally specify a unique microscopic state. Different density operators can give exactly the same measured values. The preparation procedure, however, usually restricts which microscopic states are physically possible. If \(\mathfrak P\) denotes this preparation class, the states compatible with a measured record are
\begin{equation}
\mathcal A_{\mathcal R,\mathfrak P}(\mathbf m)
=
\left\{
\rho\in\mathfrak P:
\Tr(\rho A_\mu)=m_\mu
\ \text{for all }\mu
\right\}.
\label{eq:compatibility}
\end{equation}
Thus one macroscopic state generally represents a whole set of microscopic states, as shown in Fig.~\figpanel{fig:framework}{b}. The practical question is whether the differences inside this set matter for the quantity that we want to predict.

What counts as a sufficient thermodynamic state therefore depends on the complete experimental task. We describe that task by
\begin{equation}
\Theta_{\rm eng}
=
\left(
B,\mathcal R,\mathfrak P,\mathcal C,\mathcal E,\mathfrak M,\mathcal T
\right),
\label{eq:theta_eng}
\end{equation}
where \(B\) specifies the physical boundary, \(\mathcal R\) the retained record, \(\mathfrak P\) the preparation domain, \(\mathcal C\) the available controls, \(\mathcal E\) the environment and allowed resources, \(\mathfrak M\) the measurement architecture, and \(\mathcal T\) the task.
Here $\mathcal R$ denotes a candidate record, not a record already chosen for every task. We hold the physical resources and prediction requirements fixed while comparing accessible candidates. There is therefore no universal smallest thermodynamic state: changing the task, controls, preparation, or measurement scheme can change which variables are worth retaining.

\subsection{What does the retained record determine exactly?}
\label{subsec:exact_closure}

Before introducing any approximation, we first ask what the measured record already determines exactly. The retained observables generate the real operator space
\begin{equation}
\mathcal V_{\mathcal R}
=
\operatorname{span}_{\mathbb R}
\left\{
\mathbb I,A_1,\ldots,A_m
\right\}.
\label{eq:record_span}
\end{equation}
Over the full quantum state space, a Hermitian observable \(X\) is fixed by the retained record if and only if \(X\in\mathcal V_{\mathcal R}\). A restricted preparation class can reduce the remaining freedom further, so an observable that is not globally fixed may still be fixed on the physically allowed preparation domain. This condition also determines when work and specified future observables are known from the record (see Note~\ref{supp:note-closure} of the SM \cite{SuppMat} for the exact theorem and its proof).

The simplest example is work metering. If the Hamiltonian contains a controlled parameter \(\lambda_a\), the corresponding generalized force is \(\partial H/\partial\lambda_a\). The instantaneous work associated with this control can therefore be obtained exactly from the measured thermodynamic
state whenever \(\partial H/\partial\lambda_a\in\mathcal V_{\mathcal R}\). A small macroscopic record can therefore determine the work rate of a given drive even when it leaves most of the microscopic state unknown. Work metering is a narrower task than full state reconstruction.

The same reasoning applies to future prediction. Suppose a prescribed control protocol \(u\) generates the unitary propagator \(U_u\), and we want to predict the expectation value of an observable \(O\) at the end of the protocol. From
the state at the initial time, \(\langle O\rangle_{\rm f}=
\Tr[\rho\,U_u^\dagger O U_u]\). The future observable is therefore determined exactly by the input thermodynamic record whenever the pulled-back operator \(U_u^\dagger O U_u\) lies in \(\mathcal V_{\mathcal R}\).

Finite-protocol work has the same structure. If \(H_i\) and \(H_f\) are the Hamiltonians at the beginning and end of the protocol, respectively, the mean work done on the system can be written as
\begin{equation}
W_{\rm on}[\rho,u]
=
\Tr[\rho X_W(u)],
\quad
X_W(u)
=
U_u^\dagger H_f U_u-H_i.
\label{eq:work_functional}
\end{equation}
The retained state predicts this work exactly whenever
\(X_W(u)\in\mathcal V_{\mathcal R}\). Instantaneous work metering and finite-time prediction are therefore two forms of the same operator-closure problem.

\subsection{How much information is missing?}
\label{subsec:ambiguity}

Exact closure will not hold for every observable. When it fails, we need a quantitative measure of what remains unresolved rather than simply declaring the reduced state incomplete.

For a target observable \(X\), we define its distance from the retained operator space as
\begin{equation}
d_{\mathcal R}(X)
=
\inf_{A\in\mathcal V_{\mathcal R}}
\|X-A\|_\infty .
\label{eq:operator_distance}
\end{equation}
This distance has a direct operational meaning. Since the identity belongs to \(\mathcal V_{\mathcal R}\), the largest possible difference in the expectation of \(X\) between two microscopic states that give the same measured record is exactly
\begin{equation}
\sup_{\substack{\rho,\sigma\\
\Tr[(\rho-\sigma)A_\mu]=0\ \forall\mu}}
\left|
\Tr[X(\rho-\sigma)]
\right|
=
2d_{\mathcal R}(X).
\label{eq:global_ambiguity}
\end{equation}
Thus \(d_{\mathcal R}(X)\) measures how strongly the requested observable can respond to microscopic differences invisible to the record. This is a finite-dimensional observability statement, with a classical counterpart for commuting observables; the specifically quantum features enter through noncommuting operators and their measurement costs (see Note~\ref{supp:note-ambiguity} of the SM \cite{SuppMat} for the duality and preparation-dependent bounds).

The same geometry applies when the quantity of interest is a coarse probability distribution rather than a single expectation value. If \(\{P_r\}\) is a projective coarse measurement, any binary event \(S\) is represented by \(P_S=\sum_{r\in S}P_r\). Applying Eq.~\eqref{eq:global_ambiguity} to these event projectors gives the corresponding total-variation ambiguity. Work prediction and future macroscopic-state prediction can therefore be treated within the same framework.

Equation~\eqref{eq:global_ambiguity} is deliberately conservative: it allows every density operator with the same retained record. A calibrated preparation procedure can restrict that set substantially and reduce the physically relevant ambiguity. The preparation-restricted question is therefore different from global ambiguity (see Note~\ref{supp:note-ambiguity} of the SM \cite{SuppMat} for the corresponding certificates).

\subsection{Closing the dynamics over a finite horizon}
\label{subsec:state_history}

The previous results tell us whether a given record is sufficient, but they do not yet tell us what additional information should be retained when it is not. For a finite prediction horizon, there are two natural ways to supply the
missing dynamics.

The first is to enlarge the instantaneous state. Consider a projective record with probabilities \(p_r(t)=\Tr[\rho(t)P_r]\). For the two driven records studied below, the explicit \(i[\dot H,P_r]\) term in the general second derivative vanishes, so the first two local dynamical layers are
\begin{equation}
D_r^{(1)} = i[H,P_r],
\quad
D_r^{(2)} = -[H,[H,P_r]].
\label{eq:derivative_layers}
\end{equation}
The expectation of \(D_r^{(1)}\) gives the instantaneous current of the coarse probability \(p_r\), while \(D_r^{(2)}\) gives its next local dynamical layer. Measuring such observables therefore adds direct information about how the present macroscopic state is moving. This is the local engineering form of the hidden-current structure identified in Ref.~\cite{Ahmadi2026Endogenous}:
when the present record is incomplete, the missing microscopic structure can carry the first dynamical directions that the record itself does not contain.

The second option is to retain earlier measurements of the same coarse record. Those values contain information about how the state reached the present time. In the measurement benchmark used here, they are stored ensemble estimates obtained from separately prepared copies, not repeated undisturbed projective measurements of one quantum realization. For a sufficiently short lag \(s\),
\begin{equation}
p_r(t_c-s)
=
p_r(t_c)
-
s\,\dot p_r(t_c)
+
\frac{s^2}{2}\ddot p_r(t_c)
+
\mathcal O(s^3).
\label{eq:history_expansion}
\end{equation}
When the lag is very short, the difference between past and present estimates measures the local current. A second past point can also estimate curvature, but subtracting nearly equal averages amplifies statistical noise. At longer lags, the back-transported record can carry operator directions outside the first two local derivative layers. This does not place history outside the complete derivative hierarchy; it means that a short, truncated instantaneous description does not capture all of its useful finite-time information.

We thus have two classical-looking ways to represent missing dynamics: enlarge the present state or supplement a smaller state with history~\cite{ColemanGurtin1967,Day1972,Chorin2002,Rupe2022}. They need not be equivalent and can require very different numbers of measurements. Here ``memory'' means predictive information stored in the earlier records. It does not imply coupling to an external bath. Choosing between them is therefore part of the thermodynamic design problem, as illustrated in Fig.~\figpanel{fig:framework}{d}
\cite{Grabert1982,LeBellac2004,Lebon2008}. The two descriptions are related by the exact dynamics (see Note~\ref{supp:note-dynamics} of the SM \cite{SuppMat} for back-transported history operators).

\subsection{Constitutive accuracy and measurement cost}
\label{subsec:task_budget}

A reduced description is useful only if its intrinsic prediction error and its measurement uncertainty together remain below the required task tolerance. This means that constitutive closure and metrology cannot be optimized independently.

Suppose the preparation is known to lie within trace distance
\(\epsilon_{\rm prep}\) of a nominal state, and let \(X_k\) denote one of the operators required by the prediction task. We approximate it using an operator \(A_k\in\mathcal V_{\mathcal R}\) built from the retained coordinates. A sufficient finite-confidence error certificate can be written as
\begin{equation}
\Gamma_k = \delta_{{\rm lift},k} + \kappa\|X_k-A_k\|_\infty
+ \sum_s \frac{b_{ks}(A_k)}{\sqrt{N_s}},
\label{eq:task_budget}
\end{equation}
where \(\kappa=\min(2,4\epsilon_{\rm prep})\). The first term accounts for any interpolation or continuous-time lifting used by the constitutive model. The second term bounds the remaining ambiguity inside the preparation domain. The last term is the finite-confidence measurement uncertainty, where \(N_s\) is the number of repetitions assigned to measurement setting \(s\).
All three error terms must be paid from the same task tolerance (see Note~\ref{supp:note-metrology} of the SM \cite{SuppMat} for the concentration bound and optimized shot allocation).

The important point is that reducing the constitutive residual
\(\|X_k-A_k\|_\infty\) is not the only objective. The coefficients used in \(A_k\) also determine how measurement noise is amplified and therefore how many experimental repetitions are required. We consequently choose the constitutive representation and the measurement allocation together by minimizing the total number of repetitions \(\sum_sN_s\), subject to \(\Gamma_k\le\epsilon_{\rm task}\) for every quantity required by the task.

A slightly less accurate operator approximation can therefore be the better choice if it is much easier to measure. This is a quantum version of the familiar experimental-design tradeoff between the information gained and the cost of obtaining it~\cite{ZhangAyoubSundaram2017}. This is the reason that the thermodynamic state cannot be selected from microscopic closure alone. It must be selected together with the prediction task and the metrology.

Finally, the logical meaning of the certificate is important. If a sufficient upper bound lies below the task tolerance, the chosen record is certified as adequate. If that upper bound lies above the tolerance, this alone does not prove that the record is impossible to use; a tighter treatment of the restricted microscopic compatibility set may still succeed. The positivity of admissible states may allow a tighter calculation (see Note~\ref{supp:note-ambiguity} of the SM \cite{SuppMat} for the restricted-fiber refinement). We keep this distinction throughout the numerical results below.

\section{A quantum constitutive data sheet}
\label{sec:bh}

We now test whether the framework can be assembled into an object that could actually be used like an engineering response table: a small set of quantities measured online, a constitutive response calibrated beforehand, a declared operating range, and an accuracy certificate.

The working medium is the same finite Bose--Hubbard platform used in Refs.~\cite{Ahmadi2026FiniteParticle,Ahmadi2026PhysicalWork}. There, the spatial record together with the mean energy was taken as the accessible thermodynamic information and was shown to support a complete cycle without tomography. Here the role of the model is different. We no longer ask how much irreversibility remains for a prescribed record. We ask whether microscopic dynamics can be compressed into a reusable constitutive data sheet and, before doing so, which measurable state should be retained for finite-horizon prediction.

This makes the test stronger than reproducing one trajectory. A useful data sheet must keep one small online state and one compact response over a declared parameter region. It must also predict trajectories excluded from calibration. Only then is it more than a fit to the trajectory used to build it.

\subsection{Working medium and prediction task}
\label{subsec:bh-system}
%
\begin{figure}[t]
\centering
\includegraphics[width=\linewidth]{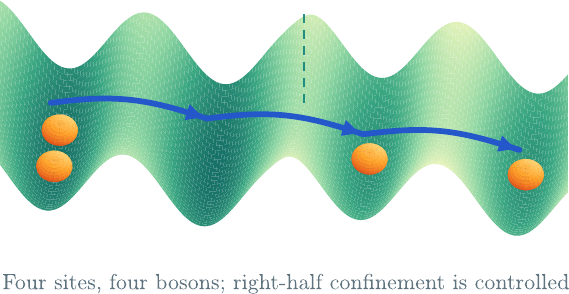}
\caption{\textbf{Four-site Bose--Hubbard working medium.}
Four bosons occupy four lattice sites, with nearest-neighbor hopping $J$ (blue arrows) and on-site interaction $U$. The controlled energy offset $\Lambda(t)$ couples to the right-half particle number $N_R$. The measured record $R(t)$ gives the probabilities of finding different numbers of bosons in that half.}
\label{fig:FQET_Bose_Hubbard_schematic}
\end{figure}

We consider four bosons on a four-site open chain. The Hamiltonian is (as illustrated in Fig. \ref{fig:FQET_Bose_Hubbard_schematic})
\begin{equation}
H(J,U,\Lambda)
=
-JT+\frac{U}{2}D+\Lambda N_R ,
\label{eq:BH}
\end{equation}
where \(J\) is the hopping energy, \(U\) is the on-site interaction strength, and \(\Lambda\) controls an energy offset between the two halves of the chain.
Here \(T=\sum_j(a_j^\dagger a_{j+1}+a_{j+1}^\dagger a_j)\) is the
nearest-neighbor hopping operator, \(D=\sum_j n_j(n_j-1)\) measures double occupation, and \(N_R\) counts the
number of bosons in the right half of the chain. The fixed-particle Hilbert space has dimension \(35\).

The macroscopic spatial record is deliberately much smaller. Let \(P_r\) project onto the states with \(r\) bosons in the right half, where \(r=0,\ldots,4\). The measured probabilities
\(p_r(t)=\Tr[\rho(t)P_r]\) form the five-outcome record \(R(t)\).
Only four of these probabilities are independent because they sum to one.

This record has a direct thermodynamic role. Since
\(N_R=\sum_r rP_r\), measuring \(R\) determines
\(\langle N_R\rangle\). The generalized force associated with the controlled confinement is therefore already contained in the coarse spatial record, because \(\partial H/\partial\Lambda=N_R\). The instantaneous work transferred by changing \(\Lambda\) can consequently be metered without reconstructing \(\rho\).

The nominal preparation is canonical for the initial Hamiltonian at \(\kb T_h=5J\). We use \(U_0/J=1.3/3\) and an initial confinement \(\Lambda_{0,0}=20J\). The system is then thermally isolated and \(\Lambda\) is reduced linearly from \(\Lambda_0\) to zero over a stroke of duration \(\tau_{BC}\), with nominal value \(J\tau_{BC,0}=20\). Because this stroke is isolated, its energy change is work; no heat crosses the system boundary.

Our thermodynamic prediction is made early in this driven evolution. At the decision time \(Jt_c=0.2\), we ask for the future spatial record \(R(t_c+\tau)\) over the whole range \(0\le J\tau\le0.2\), as illustrated in Fig.~\figpanel{fig:BHmain}{a}. The preparation is allowed to vary within a trace-distance radius \(\epsilon_{\rm prep}=0.01\) around the nominal preparation. The target is a pairwise total-variation prediction width no
larger than \(1\%\), with \(99\%\) confidence under the common ideal repeated-projective measurement model introduced above. Let \(\epsilon_{\rm TV}\) denote a uniform total-variation bound over the forecast interval. Because the right-half particle number runs from zero to four, this also bounds the difference in the mean driven force by \(4\epsilon_{\rm TV}\). The corresponding work difference during the forecast can therefore be bounded without predicting any additional observable (see Note~\ref{supp:note-bh-microscopic} of the SM \cite{SuppMat} for the work bound).

These choices define the task before any reduced state is selected. The next question is therefore operational: what should the experimenter measure at \(t_c\) in order to meet this target at the lowest measurement cost? These quantities are computed from one fixed-particle microscopic model (see Note~\ref{supp:note-bh-microscopic} of the SM \cite{SuppMat} for the basis, projectors, ramp, and propagation).

\subsection{Selecting the online thermodynamic state}
\label{subsec:bh-state}

The present spatial record does not determine its own future because it does
not contain the microscopic flow between the five record sectors. The missing
dynamics can be supplied either by measuring the current and higher local
derivative layers directly, or by retaining a small amount of recent history.
We therefore compare these two possibilities using the same prediction target,
preparation domain, confidence level, and shared-repetition optimization.
The comparison must be made before the state is fixed. Accurate history-based
prediction alone would not show that past measurements are preferable to
measuring the missing dynamics directly.

Figure~\figpanel{fig:BHmain}{c} shows the resulting task-aware comparison.
Over the tested finite-horizon range, the optimized history representation
requires fewer shared repetitions than the best instantaneous derivative
representation. The difference grows with the forecast horizon. At
\(J\tau=0.20\), the best derivative representation requires about \(8.5\)
times as many repetitions as the best history representation.

For the data sheet we need one fixed state that can be used throughout the
declared horizon rather than a different measurement architecture at every
future time. The resulting practical thermodynamic state is
\begin{equation}
\mathcal R_{\rm BH}(t_c)
=
\left\{
E(t_c),
R(t_c),
R(t_c-0.1/J),
R(t_c-0.2/J)
\right\}.
\label{eq:BHrecord}
\end{equation}
Thus the experimenter retains the mean energy, the present spatial record,
and two previous measurements of the same coarse record. No microscopic state
reconstruction is required. Since \(Jt_c=0.2\), the oldest retained record,
\(R(t_c-0.2/J)\), is the record at the beginning of the driven protocol,
\(t=0\).

The two history nodes are not arbitrary convenient choices. They emerge from
the task-aware search over temporal measurement architectures. At shorter
prediction horizons the richer history can collapse onto a smaller active
support, but the fixed two-lag architecture is sufficient for the complete
range \(0\le J\tau\le0.2\). We therefore freeze this state before building
the constitutive data sheet. This fixes one architecture before calibration (see Note~\ref{supp:note-bh-selection} of the SM \cite{SuppMat} for the history-versus-derivative comparison and support boundaries).

\subsection{From microscopic calibration to a continuous response}
\label{subsec:bh-datasheet}

A useful data sheet should remain valid when the device is not operated at
one perfectly known parameter point. We therefore allow the interaction
strength, initial confinement, and stroke duration to vary simultaneously.
The declared operating region is
\(U/U_0\in[0.9,1.1]\),
\(\Lambda_0/\Lambda_{0,0}\in[0.9,1.1]\), and
\(\tau_{BC}/\tau_{BC,0}\in[0.9,1.1]\).

Calibrating an independent response model at every point in this
three-dimensional region would defeat the purpose of a compact constitutive
description. We instead use only the eight corners of the parameter cube to
construct the response tables, as shown in
Fig.~\figpanel{fig:BHmain}{b}. The coefficients that map the measured
thermodynamic state in Eq.~\eqref{eq:BHrecord} to future coarse probabilities
are then interpolated trilinearly between these eight corners.

Using only eight response tables raises an obvious question: what prevents a
large interpolation error from appearing between the corners? We therefore
separate \emph{response calibration} from \emph{response validation}. The
eight corners define the constitutive table, while a separate
\(5\times5\times5\) lattice of \(125\) parameter points, together with
additional deterministic interior checks, is used to test how rapidly the
response changes inside the cube. These validation points are not added to
the constitutive table. Their role is to test the interpolation and estimate a conservative envelope for its parameter sensitivity. A further \(20\%\) inflation is applied to that numerically validated envelope. This provides a conditional engineering certificate, not an interval-arithmetic proof of the global maximum.

This distinction is important experimentally. A dense offline validation
campaign does not imply that the online constitutive model must itself be
dense. Once the data sheet has been constructed, the experimenter still uses
only the state in Eq.~\eqref{eq:BHrecord} and the eight-corner interpolated
response.

Under this numerical sensitivity envelope, the reported worst intrinsic upper certificate over the parameter box and forecast horizon is \(0.7304\%\). This lies below the \(1\%\) task tolerance before finite
measurement uncertainty is added. After including the common conservative
\(99\%\) repeated-projective metrology model, the complete task can be
supported with a shared benchmark of approximately
\(3.786\times10^7\) repetitions.

The latter number should not be confused with the nominal comparison in
Fig.~\figpanel{fig:BHmain}{c}. Panel (c) compares competing state
representations at the nominal operating point. The larger
\(3.786\times10^7\) benchmark refers to one common measurement allocation
that supports the complete continuous-domain data-sheet specification.

The result is therefore more than a successful reduced-state prediction at
one parameter point. It specifies a fixed measurable state and response, together with a preparation class, operating domain, horizon, tolerance, confidence level, and metrology requirement. The domain claim is conditional on the tested sensitivity envelope, as detailed in the SM. The validity of the continuous-domain result rests on the sensitivity audit (see Note~\ref{supp:note-bh-datasheet} of the SM \cite{SuppMat} for its construction and numerical status). What
remains is the most important test of such a data sheet: whether it still
works after all of those choices have been frozen.

\subsection{Blind prediction with a frozen data sheet}
\label{subsec:bh-blind}

A response table can agree with its calibration data and still fail elsewhere. The next test is therefore whether it predicts cases that were never used to construct it. We therefore froze the state in
Eq.~\eqref{eq:BHrecord}, the eight corner response tables, the trilinear
interpolation rule, and the sensitivity envelope before performing the final
validation. No coefficients were refitted, no history times were moved, and
the error envelope was not enlarged after inspecting the holdout results.

The frozen data sheet was then given to an independently implemented
microscopic solver. This second calculation reconstructs the fixed-particle
Bose--Hubbard sector independently and propagates the dynamics with a
fourth-order two-node Gauss--Magnus method, rather than the DOP853 integrator
used during calibration. The purpose of this change is to prevent agreement
from being tied to one microscopic implementation.

The holdout contains \(64\) independently scrambled Sobol points inside the
parameter cube and \(8\) additional points chosen close to its corners. At
each of the resulting \(72\) parameter points, we test the nominal canonical
preparation together with three independent positive perturbations near the
edge of the allowed preparation tube. Each preparation is followed over
\(81\) future times and tested against all \(15\) independent binary events
of the five-outcome spatial record. In total, the frozen constitutive model
makes \(349{,}920\) blind event predictions.

\begin{figure}[t]
\centering
\includegraphics[width=\linewidth]{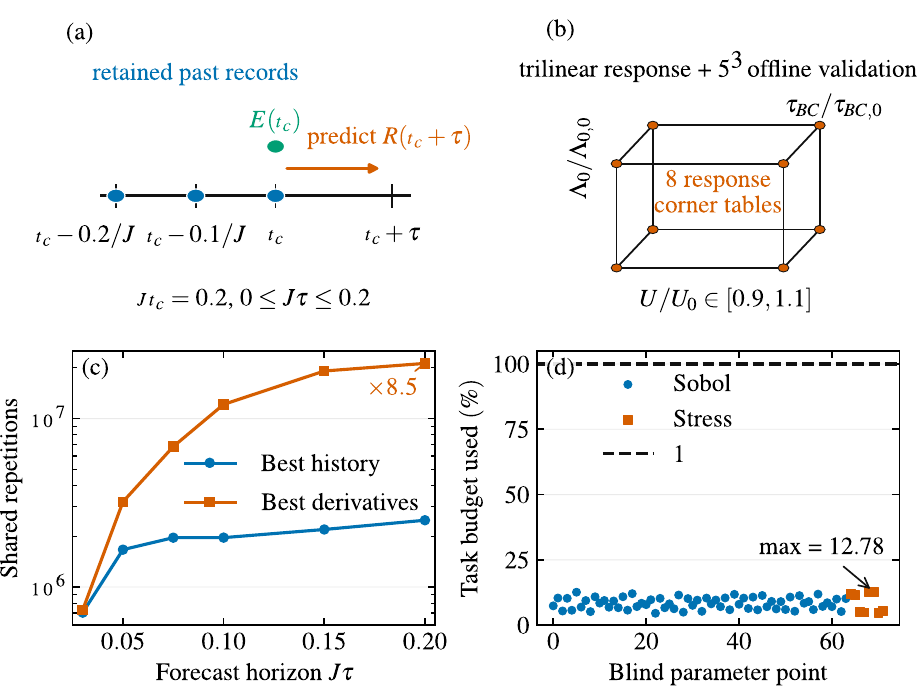}
\caption{\textbf{A quantum constitutive data sheet for the Bose--Hubbard working medium.}
\textbf{(a)} At the decision time, the online state uses present energy and the spatial record together with two ensemble records stored from earlier times to predict future right-half occupations.
\textbf{(b)} Only eight corner tables are calibrated across the \(\pm10\%\) parameter box. A separate \(5^3\) lattice tests sensitivity between corners; it does not increase the online measurement state.
\textbf{(c)} At the nominal point, finite history needs fewer repeated preparations than the best direct-derivative state. Its advantage grows to approximately \(8.5\)-fold at \(J\tau=0.20\), showing that more direct dynamical information is not always cheaper to measure.
\textbf{(d)} Independent numerical holdout at \(72\) unseen parameter points. All errors stay below the declared \(1\%\) tolerance; the largest observed error uses \(12.78\%\) of that budget. The holdout is distinct from the continuous-domain certificate.}
\label{fig:BHmain}
\end{figure}

In this independent numerical holdout, every prediction remains below the declared \(1\%\) task tolerance. The largest observed absolute event-probability error is
\(0.1278\%\), so the worst holdout case uses only \(12.78\%\) of the allowed
task budget, as shown in Fig.~\figpanel{fig:BHmain}{d}. Independent
step-size and unitarity checks place the numerical propagation error far below
this prediction scale.

The blind error and the continuous-domain certificate quantify different
things and should not be identified. The value \(0.1278\%\) is the largest
error actually observed on a finite set of unseen microscopic tests. The
\(0.7304\%\) value is a conservative sufficient upper certificate that also
accounts for unresolved microscopic variation inside the declared preparation
domain and for interpolation throughout the continuous parameter box. The
blind campaign therefore validates the frozen constitutive model; it does not
replace the certificate.

Taken together, these results realize the intended thermodynamic workflow.
Microscopic quantum dynamics is used offline to select the state, calibrate
the constitutive response, and certify where it can be used. Once this has
been done, future macroscopic behavior can be predicted online from the small
measured thermodynamic state without reconstructing the density operator. Independent propagation checks support this distinction (see Note~\ref{supp:note-bh-blind} of the SM \cite{SuppMat} for the full holdout and Solver-B convergence tests).

\section{Selecting the thermodynamic representation}
\label{sec:selection}

The Bose--Hubbard data sheet answers one part of the engineering problem: a compact finite-history state can support certified prediction over a nonzero operating domain. It does not answer whether history is the right choice in general. The same missing dynamical information could instead be measured directly through currents or higher time derivatives.

We therefore separate the success of one representation from the rule used to choose it. Adding observables enlarges the available operator space and can improve intrinsic closure, but every added coordinate must also be measured. It may require more settings, more repetitions, or a poorly conditioned combination of noisy estimates. The useful thermodynamic state is therefore the one that meets the complete prediction task at acceptable experimental cost, not simply the one that contains the most information.

\subsection{Competing descriptions of the missing dynamics}
\label{subsec:representation-families}

We compare two ways of supplementing the present energy and coarse record.
The first adds instantaneous dynamical information. We denote by
\(\mathcal J\) the independent current coordinates obtained from the first
record derivatives, and by \(K^{(2)}\) the corresponding independent
second-derivative coordinates. The two instantaneous candidate states are
therefore \((E,R,\mathcal J)\) and
\((E,R,\mathcal J,K^{(2)})\).

The second description retains past values of the coarse record. We denote by
\(H_2\) the state containing one past snapshot and by \(H_3\) the state
containing two,
\begin{equation}
\begin{aligned}
H_2 &=
\{E(t_c),R(t_c),R(t_c-s_1)\},\\
H_3 &=
\{E(t_c),R(t_c),R(t_c-s_1),R(t_c-s_2)\}.
\label{eq:history_families}
\end{aligned}
\end{equation}
The lags \(s_1\) and \(s_2\) are not fixed in advance. They are part of the
representation design and are chosen for the declared prediction task.

This construction also gives a natural comparison of information size. In
the five-outcome records used here, \(R\) contains four independent
probabilities. Thus \(H_2\) and \((E,R,\mathcal J)\) each contain nine
independent scalar coordinates, while \(H_3\) and
\((E,R,\mathcal J,K^{(2)})\) each contain thirteen. Equal coordinate count
does not imply equal laboratory cost, but it prevents a trivial comparison in
which one family is allowed simply to retain much more macroscopic
information than the other.

The two representations probe motion in different ways. A short past record can approximate a current through a finite difference, while a longer lag can carry trajectory information beyond the retained current and curvature layers. We therefore do not assume that a history
coordinate is merely an indirect derivative measurement. Whether that is true
must be determined from the dynamics; the state--history relation can be tested directly (see Note~\ref{supp:note-dynamics} of the SM \cite{SuppMat} for its operator form).

\subsection{The most accurate constitutive fit need not be the cheapest}
\label{subsec:cost-aware-selection}

For a fixed candidate state, many constitutive coefficient sets can predict
the same future observable with different compromises between intrinsic error
and measurement noise. Choosing the coefficients that minimize only the
operator residual is therefore not, in general, the same as choosing the
coefficients that minimize the experimental cost.

The physical reason is that a constitutive prediction combines noisy measured averages. Large coefficients of opposite sign can cancel in the predicted value while adding their measurement uncertainties. A fit with an excellent microscopic residual can therefore be expensive to use. A slightly less accurate
fit can then require far fewer experimental repetitions.

We therefore use the common task budget introduced in
Eq.~\eqref{eq:task_budget}. For every required future event and time, the
unresolved-state contribution and the finite-confidence measurement
contribution must together remain below the same tolerance. The constitutive
coefficients, history lags, and shared measurement allocation are optimized
jointly to minimize the required experimental repetitions.

This changes the meaning of ``best reduced model.'' The goal is not the
smallest constitutive residual. It is the least costly measured
representation that still satisfies the declared thermodynamic task. The resulting cost separates the remaining accuracy margin from the statistical burden (see Note~\ref{supp:note-metrology} of the SM \cite{SuppMat} for the shared-shot optimization and scalar factorization).

\subsection{A richer candidate state need not be an active state}
\label{subsec:nested-supports}

There is a second issue that becomes important when comparing representations
of different size. The candidate families are nested:
\((E,R,\mathcal J)\subset(E,R,\mathcal J,K^{(2)})\) and
\(H_2\subset H_3\). A richer candidate state can therefore reproduce the
smaller one simply by assigning zero constitutive weight to its additional
coordinates.

For this reason, the appearance of a variable in the candidate state does not
mean that the experimenter should actually measure it. For example, a calculation performed in the \(H_3\) family may assign no useful weight to the second historical record. The effective
state is then \(H_2\), even though the optimization was allowed to search the
larger \(H_3\) space.

We therefore distinguish a \emph{candidate support}, which contains every variable offered to the optimizer, from an \emph{active support}, whose extra variables are actually worth measuring. Before claiming that an additional coordinate is
useful, we compare the larger and smaller states in both directions. The
optimized smaller-state solution is embedded into the larger family with zero
weight on the new coordinates. Conversely, a larger-state solution is
projected back onto the smaller family and reoptimized. The new layer is
called active only if it acquires a genuine measurement role and lowers the
best cost obtained after these cross-seeded comparisons.

This precaution is necessary because numerical optimization can otherwise
mistake different optimization basins for a physical state transition. More
importantly, it gives the experimental statement we actually need: a new
coordinate is retained only when the improvement it provides is large enough
to justify measuring it.

\subsection{Family selection and layer activation are different questions}
\label{subsec:family-vs-support}

The nested structure separates two decisions that should not be confused.
The first is \emph{family selection}: should the missing dynamics be
represented by finite history or by direct instantaneous dynamical
observables? The second is \emph{layer activation}: once a family has been
chosen, is another history node or another derivative layer worth measuring?

These transitions need not occur at the same prediction horizon. A second
history point can become useful while the history family as a whole is still
more expensive than direct current measurement. Likewise, \(K^{(2)}\) can
become useful inside the instantaneous family without making that family
preferable to history. The thermodynamic-state problem therefore cannot be
reduced to asking when a particular operator first improves a constitutive
fit.

There is also a distinction between coordinates required by the complete
thermodynamic task and coordinates used only by the forecasting subproblem.
For example, a forecasting optimizer may place almost no additional
measurement weight on the mean energy, while energy can still be required for
work or energy accounting. We therefore keep mandatory thermodynamic
coordinates separate from optional predictive coordinates. State selection
always refers to the complete task \(\mathcal T\), not to one numerical
subproblem considered in isolation.

\subsection{The output is generally a Pareto choice}
\label{subsec:pareto-selection}

Even after the task-aware optimization, there is not always a unique
laboratory winner. One representation may require fewer repetitions but more
distinct measurement settings; another may use fewer settings but require
more repeated preparations. In a real experiment, these resources can also
differ in calibration effort, latency, detector complexity, disturbance, and
hardware requirements.

Without a specific experimental platform, assigning all of these costs one
universal numerical weight would be arbitrary. We therefore use the number of
shared repetitions as a common quantitative benchmark while keeping setting
count and representation complexity visible as separate resources. When one
state is better in every reported resource, it is clearly preferred. When
different resources favor different states, the correct result is a Pareto
tradeoff rather than an artificial single winner.

The Bose--Hubbard system lies in a regime where finite history remains
strongly competitive under this task-aware comparison. The larger state should therefore be selected only when the added measurement has an active role (see Note~\ref{supp:note-support} of the SM \cite{SuppMat} for the cross-seeding and Pareto criteria). That result alone does
not establish a universal preference for memory. To see whether the selection
procedure itself is general, we now apply the same logic to a microscopically
different system in which the preferred representation changes with the
forecast horizon.

\section{A contrasting Ising regime}
\label{sec:ising}
%
\begin{figure}[t]
\centering
\includegraphics[width=\linewidth]{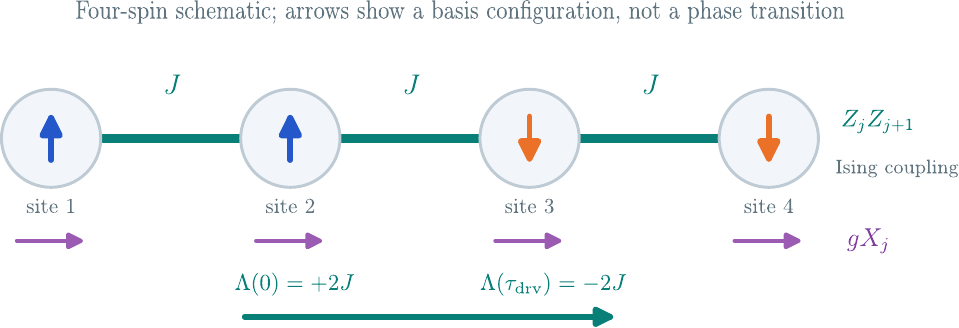}
\caption{\textbf{Driven transverse-field Ising chain.}
Four spins interact through nearest-neighbor coupling $J$ and evolve under a transverse field $g$. The longitudinal control $\Lambda(t)$ couples to the total number of up spins $N_\uparrow$. The measured record $R(t)$ gives the probabilities of different up-spin counts. The arrows illustrate a possible spin configuration.}
\label{fig:FQET_Ising_chain_schematic}
\end{figure}

The Bose--Hubbard result would be much less informative if the same type of state were always preferred. We therefore apply the selection procedure to a microscopically different working medium and ask whether it can identify a different representation when the dynamics demands one.
We use a driven transverse-field Ising chain (as illustrated in Fig. \ref{fig:FQET_Ising_chain_schematic}). The model is rebuilt independently of the Bose--Hubbard calculation: no history times, derivative scales, or preferred state architecture are transferred from the first system. The comparison therefore tests the state-selection principle rather than the portability of one fitted representation.

\subsection{Working medium and thermodynamic record}
\label{subsec:ising-system}

The working medium is an open chain of four spins with Hamiltonian
\begin{equation}
H(t)
=
-JC_{zz}
-gX_{\rm tot}
+\Lambda(t)N_{\uparrow}.
\label{eq:ising-H}
\end{equation}
Here \(J\) sets the nearest-neighbor interaction strength,
\(C_{zz}=\sum_{j=1}^{L-1}Z_jZ_{j+1}\), \(g\) is the transverse-field
strength, and \(X_{\rm tot}=\sum_jX_j\). The controlled parameter
\(\Lambda(t)\) couples to the total number of up spins,
\(N_{\uparrow}=\sum_j(\mathbb I+Z_j)/2\).

The initial state is canonical at \(k_{\rm B}T=2J\). During the protocol the
system is thermally isolated and \(\Lambda\) is ramped linearly from \(+2J\)
to \(-2J\) over \(J\tau_{\rm drv}=4\). The transverse field \(g\) is varied
between the different operating points discussed below.

The coarse thermodynamic record is the probability distribution of the total
up-spin number. If \(P_r\) projects onto the sector containing \(r\) up spins,
then \(R(t)\) consists of the five probabilities
\(p_r(t)=\Tr[\rho(t)P_r]\), with \(r=0,\ldots,4\). As in the Bose--Hubbard
case, only four of them are independent.

This choice again has a direct thermodynamic meaning. Since
\(N_{\uparrow}=\sum_r rP_r\), the coarse record determines
\(\langle N_{\uparrow}\rangle\) and therefore closes the generalized force
associated with the controlled parameter \(\Lambda\). The same record can
thus be used for exact online metering of the work transferred through the
\(\Lambda\) drive.

For future-record prediction, however, static work closure is not enough.
Mean energy supplies one combination of
\(\langle C_{zz}\rangle\) and \(\langle X_{\rm tot}\rangle\), and adding
\(C_{zz}\) closes the remaining static generalized forces when \(g\neq0\).
Yet this additional static variable gives almost no improvement in the
short-time prediction of \(R\). This provides a useful distinction: the
variables needed to meter the controls need not be the variables that best
predict the future macroscopic state.

We therefore turn directly to dynamically aligned information. At the
decision time \(Jt_c=1\), we compare the history family
\(H_2\rightarrow H_3\) with the instantaneous family
\((E,R,\mathcal J)\rightarrow(E,R,\mathcal J,K^{(2)})\), as shown in
Fig.~\figpanel{fig:ising}{a}. The preparation radius is
\(\epsilon_{\rm prep}=0.01\), the target pairwise total-variation width is
\(1\%\), and the common metrology benchmark uses \(99\%\) family-wise
confidence. The two families are evaluated with the same task budget (see Note~\ref{supp:note-ising} of the SM \cite{SuppMat} for the model, time lift, and representation map).

\subsection{Three different state-selection boundaries}
\label{subsec:ising-phase}

We now ask how the preferred thermodynamic representation changes as the
required forecast horizon grows. This requires three separate questions.

First, while both families use only their first dynamical layer, which one is
cheaper to measure? This defines the family boundary
\(\tau_{\rm family}\), where the optimized repetition counts of \(H_2\) and
\((E,R,\mathcal J)\) are equal.

Second, at what horizon does one past record cease to be enough inside the
history family? This defines the support transition \(H_2\rightarrow H_3\)
at \(\tau_{H_3}\).

Third, at what horizon does the direct second-derivative layer become worth
measuring inside the instantaneous family? This defines the transition
\((E,R,\mathcal J)\rightarrow(E,R,\mathcal J,K^{(2)})\) at
\(\tau_{K^{(2)}}\).

The task-aware optimization gives the same ordering at every transverse field
that we tested:
\begin{equation}
\tau_{\rm family}
<
\tau_{H_3}
<
\tau_{K^{(2)}}.
\label{eq:ising-boundary-order}
\end{equation}
Figure~\figpanel{fig:ising}{b} shows these three boundaries for
\(0.75\le g/J\le2\). The separation between them is important. The system
first changes which representation family is cheaper; only later does the
history family activate another past measurement, and later still does the
instantaneous family activate the direct second-derivative layer. These are
boundaries between preferred engineering representations, not phase
transitions of the microscopic Ising Hamiltonian.

At the nominal point \(g/J=1.5\), for example, the family crossover occurs
at \(J\tau_{\rm family}\simeq0.02921\). The second history node becomes active
at \(J\tau_{H_3}\simeq0.04108\), whereas the direct \(K^{(2)}\) layer does not
activate until \(J\tau_{K^{(2)}}\simeq0.0715\).

These are genuine support changes rather than labels assigned to nominally
larger candidate states. At the first validated point above the \(H_3\)
boundary, the newly activated history setting receives about
\(7.8\)--\(10.8\%\) of the optimized repetition budget across the tested
transverse fields. At the corresponding \(K^{(2)}\) transition, the new
second-derivative settings receive about \(11.3\)--\(19.6\%\). The additional
coordinates therefore acquire a finite experimental role when they enter.

The family boundary instead compares the two ways of obtaining the missing dynamics. Below
\(\tau_{\rm family}\), \(H_2\) requires both fewer repetitions and fewer
ideal measurement settings than the current-based state. History is therefore
preferred in both resource measures used here. Above this boundary, direct
current measurement requires fewer repetitions, while finite history still
uses fewer settings. The comparison then becomes a Pareto tradeoff rather
than a unique laboratory winner.

Importantly, activating \(H_3\) at a later horizon does not restore a
lower-repetition history phase. The direct instantaneous family remains the
lower-shot choice at the validated points beyond the history activation.
The state-selection result is therefore not simply that increasingly long
forecasts require increasingly long memory.

The locations of the boundaries also move systematically with the transverse
field. For the present model and metrology, the family crossover is organized
approximately by \(\langle g\tau_{\rm family}\rangle=0.04268\), while direct
second-derivative activation gives
\(\langle g\tau_{K^{(2)}}\rangle=0.10634\). Their coefficients of variation
over the tested field range are \(3.07\%\) and \(1.60\%\), respectively.
The corresponding history-activation quantity,
\(\langle g\tau_{H_3}\rangle=0.06532\), has a larger variation of
\(8.90\%\). These numbers are properties of the present Ising task and
measurement model, not universal constants. The larger variation of the history boundary suggests that its useful lag depends on more than one local dynamical rate. A stored past record probes a finite part of the trajectory, rather than just its slope at one instant.

\subsection{History itself has more than one form}
\label{subsec:ising-history-branches}

The phase map tells us when another historical coordinate becomes worth
measuring, but it does not yet tell us what that coordinate represents.
To answer this, we examine the internal structure of \(H_3\) at the nominal
point \(g/J=1.5\).

A search over the two history lags reveals two distinct regions. One uses
nearby past measurements, with lags close to
\((Js_1,Js_2)=(0.06,0.20)\). We call this the local branch because both
measurements remain relatively close to the decision time. A second branch
keeps the first lag near \(Js_1=0.06\) but moves the second much farther into
the past, near \(Js_2=0.70\).

The two branches use the same present record but ask the second past measurement to do different jobs: one mainly improves a local finite difference, while the other probes a more distant part of the trajectory. To expose this distinction without mixing it with the
final family-selection problem, we compare them using the residual-first
shared-shot construction as a separate mechanism study. Their measurement
costs cross near \(J\tau\simeq0.0584\), as shown in
Fig.~\figpanel{fig:ising}{c}. The optimized second lag changes strongly across
this crossing, showing that the two solutions are distinct temporal
architectures rather than one minimum moving smoothly through lag space.

This branch crossing should not be confused with
\(\tau_{\rm family}\), \(\tau_{H_3}\), or \(\tau_{K^{(2)}}\).
Those three quantities come from the final task-aware state-selection problem.
The crossing in Fig.~\figpanel{fig:ising}{c} is instead a diagnostic of the
internal structure available within the two-lag history family. Their different roles become clear after the present record is removed (see Note~\ref{supp:note-ising-internal} of the SM \cite{SuppMat} for the two branches and their geometry). In the next section we place this structure beside
the Bose--Hubbard result.

\subsection{Independent numerical audit}
\label{subsec:ising-validation}

Because the phase boundaries depend on small differences between optimized
operator distances, we also test the microscopic calculation with an
independent implementation. A second solver reconstructs the spin operators
from scratch and propagates the required transition operators with a
fourth-order Gauss--Magnus method using \(J\Delta t=5\times10^{-4}\).

\begin{figure}[t]
\centering
\includegraphics[width=\linewidth]{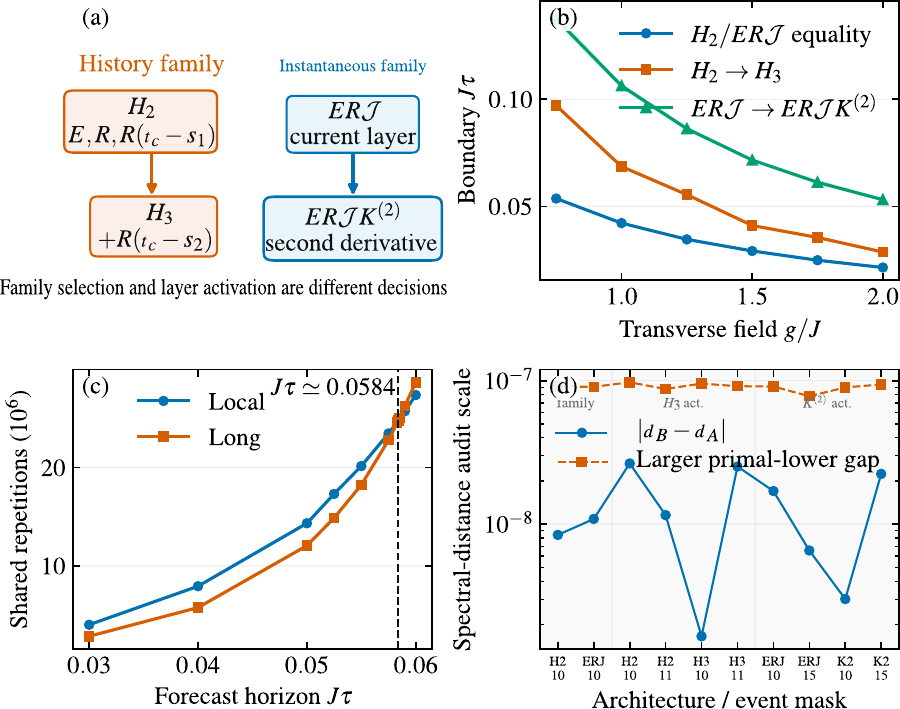}
\caption{\textbf{Choosing the measured state in the driven Ising chain.}
\textbf{(a)} Two nested choices are compared: one or two earlier records, and direct current observables with or without the second derivative. Adding a layer is distinct from choosing the cheaper family.
\textbf{(b)} As the forecast horizon grows, the repetition-cost crossover comes first, followed by activation of the second history node and then the direct curvature layer, at every tested transverse field. These are representation boundaries, not microscopic phase transitions.
\textbf{(c)} Within the two-lag history family, a nearby and a longer-memory architecture exchange order near \(J\tau\simeq0.0584\). This residual-first diagnostic is not one of the boundaries in panel (b).
\textbf{(d)} An independent propagation algorithm recomputes decisive constraints around the three nominal boundaries. Its small spectral-distance differences test the microscopic numerics, not global optimality of the outer search.}
\label{fig:ising}
\end{figure}

We audit ten selected event and architecture constraints around the nominal
\(g/J=1.5\) family crossover, \(H_3\) activation, and \(K^{(2)}\) activation.
The largest difference between the two solvers in the target operators is
\(1.133\times10^{-10}\) in operator norm, while the largest difference in the
record operators is \(6.224\times10^{-11}\). The independently calculated
spectral-distance upper values differ by at most \(2.651\times10^{-8}\), and
all certified intervals from the two implementations overlap. The numerical
audit is summarized in Fig.~\figpanel{fig:ising}{d}.

These checks validate the microscopic operator geometry on which the phase
map is built. They do not turn the outer search over history lags,
constitutive coefficients, and shared measurement resources into a
theorem-level global optimization. That outer problem is obtained from broad
screening, local refinement, continuation, and the cross-seeding procedure
described above. We therefore interpret the resulting boundaries as
well-resolved engineering optima for the declared search and task, rather
than as mathematical global-optimality theorems. The independent solver therefore checks the microscopic calculation, not the global optimum (see Note~\ref{supp:note-ising-internal} of the SM \cite{SuppMat} for the spectral and convergence audits).

The Ising system therefore gives a different answer from the Bose--Hubbard
working medium. Finite history is preferred at the shortest horizons, but
direct dynamical measurements become cheaper in repetitions as the forecast
is extended. The important result is not which representation wins in one
model. It is that the same thermodynamic framework can determine when a
different representation should be used.

\section{Why memory wins in one system and derivatives in the other}
\label{sec:mechanism}

We have now used the same state-selection method on two different working media, and it has given different answers. In the Bose--Hubbard system, finite history remains the lower-shot representation throughout the tested finite-memory window. In the Ising system, history is favored only at the shortest horizons, after which direct dynamical observables become cheaper in repetitions. The contrast lets us ask a more useful question than which representation wins: what physical property of the retained information makes one representation economical and the other costly?

A past measurement can play two different roles. If it is taken sufficiently close to the present, it mainly reproduces local derivative information through a finite difference. If it lies farther along the trajectory, it can instead provide an operator direction missing from the truncated instantaneous space of current and curvature. Figure~\figpanel{fig:mechanism}{a} illustrates these two possibilities. We call the first a local temporal jet. The second is finite-lag memory relative to the first two derivative layers; it is not information outside every higher time derivative.

To distinguish them quantitatively, we compare the operator spaces carried by the historical measurements with those carried by the direct derivative measurements. We first remove the information already present at the decision time, so that the comparison concerns only what each additional coordinate contributes. The first historical layer is then compared with the current subspace \(\mathcal J\), and the second historical layer, after removing the first, is compared with the direct \(K^{(2)}\) subspace.

We characterize this alignment using subspace overlaps and principal angles. An overlap close to one, together with small principal angles, means that the historical measurement is mainly probing the same operator directions as the corresponding direct derivative measurements. A small overlap and large angles mean that the history adds information that is poorly represented by the derivative subspace being compared. This geometric test tells us what kind of information the history contains. It does not, by itself, decide which representation is experimentally better; that also depends on the remaining constitutive margin and on measurement conditioning. The measures are defined consistently for both platforms (see Note~\ref{supp:note-mechanism} of the SM \cite{SuppMat} for principal angles, Gram conditioning, and numerical values).

\subsection{Bose--Hubbard history beyond the first two derivative layers}
\label{subsec:bh-mechanism}

We first apply this test to the Bose--Hubbard data-sheet state. Its practical
history nodes are \(Js_1=0.10\) and \(Js_2=0.20\). After removing the common
present-time information, the first historical layer has overlap \(0.3133\)
with the current subspace and a mean principal angle of \(56.03^\circ\).
The conditioned second layer has overlap \(0.2712\) with the
\(K^{(2)}\) subspace and a mean principal angle of \(58.68^\circ\).
Figure~\figpanel{fig:mechanism}{b} places these values in the wider lag scan.

These angles are far from the small-angle regime expected if the past records
were simply finite-difference implementations of the current and curvature.
To make that comparison explicit, we also evaluate a deliberately local pair,
\((Js_1,Js_2)=(0.025,0.050)\). Its corresponding overlaps are \(0.9426\)
and \(0.9148\), with mean principal angles of only \(13.81^\circ\) and
\(16.79^\circ\). The local history therefore behaves as expected: it closely
tracks the instantaneous derivative jet. The practical pair does not.

The relevant Bose--Hubbard times are also not small compared with the local
dynamical scales. Using the characteristic current and curvature rates gives
\(\Omega_1s_1\simeq2.37\) and \(\Omega_2s_2\simeq4.28\) for the practical
pair. The retained measurements therefore sample a finite part of the
trajectory rather than an infinitesimal neighborhood of the decision time.

This is why the Bose--Hubbard history is useful: its two past measurements probe task-relevant directions that are poorly represented by the current and curvature retained in our instantaneous candidate family. It is finite-time predictive information, not a claim that history lies outside the complete derivative hierarchy.

\subsection{The practical history is also better conditioned}
\label{subsec:bh-conditioning}

Geometry alone is not enough. Two coordinates can carry useful information
and still be experimentally expensive if they are nearly redundant. We
therefore also examine the Gram matrix of the normalized retained operators.
Its condition number measures how close the recorded operator directions are to linear dependence. When two measurements carry nearly the same information, extracting their small difference needs large coefficients and amplifies statistical noise.

The contrast in the Bose--Hubbard system is large. The deliberately local
history pair has a normalized Gram condition number of about
\(3.92\times10^5\), whereas the practical pair has a value of only
\(2.31\times10^3\). The practical history is therefore roughly \(170\)
times better conditioned under the same convention. This comparison is shown
together with the Ising values in Fig.~\figpanel{fig:mechanism}{c}.

This result also clarifies an otherwise surprising feature of the
Bose--Hubbard state-selection map. Direct \(K^{(2)}\) information becomes
useful inside the derivative family already near
\(J\tau\simeq0.051\)--\(0.052\), whereas the second historical node does not
become active until approximately \(J\tau=0.150\)--\(0.1525\). Nevertheless,
the history family remains cheaper in repetitions throughout the tested
finite-memory window. Early activation of a derivative layer therefore does
not mean that the derivative representation as a whole is the better
thermodynamic state.

\subsection{Ising history collapses toward the local derivative jet}
\label{subsec:ising-mechanism}

The Ising task shows the opposite pattern. As the forecast horizon grows, the optimized historical measurements move closer to the decision time. The history then becomes increasingly similar to an
indirect finite-difference measurement of the truncated local derivative space.

\begin{figure}[t]
\centering
\includegraphics[width=\linewidth]{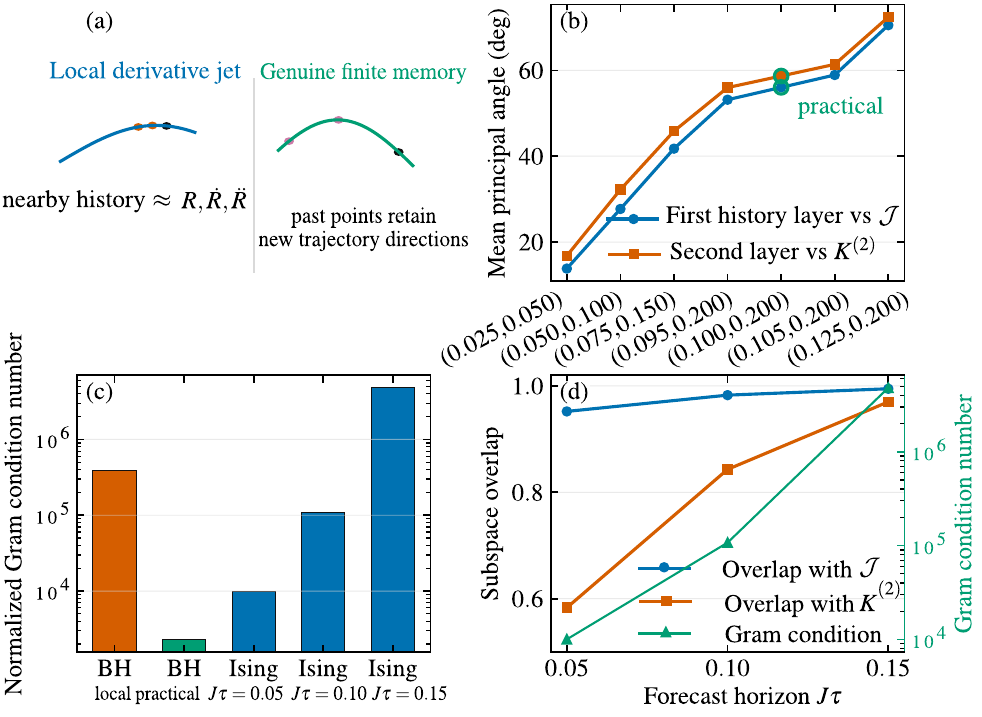}
\caption{\textbf{Why history helps in one system and costs more in the other.}
\textbf{(a)} Closely spaced records mostly recover local slopes and curvature by finite differences; longer lags can probe directions missing from that truncated derivative space.
\textbf{(b)} In the Bose--Hubbard model, the practical history has small overlaps and large principal angles relative to direct current and curvature, unlike a deliberately local lag pair.
\textbf{(c)} The practical Bose--Hubbard history is much better conditioned than local history. The Ising history instead becomes nearly redundant as its selected lags shorten.
\textbf{(d)} As the Ising forecast grows, overlap with the first two derivative layers approaches one while the Gram condition number rises. Predictive information, constitutive margin, and noise amplification must therefore be judged together.}
\label{fig:mechanism}
\end{figure}

Figure~\figpanel{fig:mechanism}{d} makes this change explicit. In the
equal-coordinate history study, at \(J\tau=0.05\) the first historical layer
has overlap \(0.952\) with the current subspace, while the conditioned second
layer has overlap \(0.583\) with \(K^{(2)}\). At \(J\tau=0.10\), these
overlaps increase to \(0.983\) and \(0.843\). By \(J\tau=0.15\), they have
reached \(0.995\) and \(0.970\). The history state has therefore moved very
close to the local temporal jet.

At the same time, its conditioning deteriorates rapidly. The normalized Gram
condition number grows from approximately \(1.00\times10^4\) at
\(J\tau=0.05\), to \(1.08\times10^5\) at \(J\tau=0.10\), and then to
\(4.80\times10^6\) at \(J\tau=0.15\). The measurements contain increasingly
similar information and must be combined through increasingly delicate
finite differences.

This explains why direct dynamical observables eventually become cheaper.
When the selected history becomes derivative-like, it loses the advantage that the Bose--Hubbard history retained. Instead of supplying
a new trajectory direction, it approximates information that can be measured
more directly through \(\mathcal J\) and \(K^{(2)}\), while paying an
additional conditioning penalty.

The two-lag Ising calculation discussed above is consistent with this picture.
It contains both a local derivative-like branch and a genuinely nonlocal
branch. Thus the statement is not that the Ising dynamics cannot contain
finite memory. Rather, under the declared forecasting task and metrology, the selected equal-coordinate history becomes increasingly derivative-like as the prediction demand changes (see Note~\ref{supp:note-ising-internal} of the SM
\cite{SuppMat}).

\subsection{A common representation-selection principle}
\label{subsec:selection-principle}

The two systems can now be understood within one physical picture. Whether a candidate thermodynamic coordinate is useful depends first on its \emph{dynamical geometry}: does it add information that is genuinely new and aligned with the required future observable, or does it largely duplicate information already available in another form?

That new information must also provide enough \emph{constitutive margin}. A representation can be easy to measure but still be unattractive if its unresolved microscopic error consumes most of the allowed prediction tolerance. Conversely, excellent intrinsic closure can be of little practical use if almost no error budget remains for the measurements.

Finally, the useful information must be experimentally accessible with acceptable \emph{metrological conditioning}. Nearly redundant coordinates can require large cancelling coefficients and many repetitions even when their combined operator span gives an accurate constitutive fit.
Representation selection is therefore controlled jointly by dynamical geometry, constitutive margin, and metrological conditioning. None of these three ingredients is a universal selector by itself.

The Bose--Hubbard system favors finite history because its practical past records remain far from the local derivative subspaces while also being well conditioned. The Ising history becomes increasingly aligned with the derivative jet and increasingly ill conditioned, so direct dynamical
measurements eventually become more economical. The difference between the two platforms is therefore not a fundamental preference for memory in one case and derivatives in the other. It is a difference in how the task-relevant dynamical information is organized and how costly that information is to measure.

This leads to the broader conclusion of the two case studies: the
thermodynamic state of a finite quantum system should not be chosen from a fixed universal list of variables. It should be selected for the working medium, operating domain, prediction horizon, task tolerance, and available metrology. The independent validation layers and geometric diagnostics are listed together (see Note~\ref{supp:note-mechanism} of the SM \cite{SuppMat} for the complete ledger).

\section{Summary and Discussion}
\label{sec:discussion}

Thermodynamics becomes useful when a few measured quantities can support reliable predictions. Classical nonequilibrium theory established the roles of constitutive laws, internal variables, and memory. Here we bring that logic to finite quantum working media and make the choice of measured information part of the task. Exact operator geometry identifies what the present record determines, while offline quantum dynamics supplies and tests the response that the record alone cannot give.

The Bose--Hubbard example turns this idea into a compact constitutive data sheet. A present spatial record is enough to meter the controlled force but not to predict its own future, so the selected state also retains two earlier records. The frozen response is tested over a declared parameter range and on independent numerical holdouts. The Ising example explains why history cannot be chosen in advance: as the required forecast changes, direct measurements of dynamical observables can use fewer repetitions. In both systems the choice depends on which operator directions are missing, how much prediction error remains, and how measurement noise is amplified.

The scope of these claims matters. Exact closure and global ambiguity are analytic results; preparation bounds are sufficient certificates. The continuous-domain data sheet relies on a numerically validated sensitivity envelope, and the outer architecture search is not a proof of global optimality. Nor do our predictive certificates replace the entropy restrictions of a general dissipative constitutive theory. A laboratory test must still account for detector noise, disturbance, latency, and calibration drift. Within these limits, the central result is practical: detailed quantum dynamics can be used beforehand so that routine thermodynamic prediction needs only a task-sufficient record and a tested response.

\section{Acknowledgments}
BA acknowledges support from IRA Program (project no. FENG.02.01-IP.05-0006/23) financed by the FENG program 2021-2027, Priority FENG.02, Measure FENG.02.01., with the support of the FNP.

\section{Data and code availability}
The numerical code used to generate the data and figures in this work is available at \url{https://github.com/Borhan19/FQET/releases/tag/v1.0-submission}. 
The repository contains the simulation scripts, plotting routines, and parameter files required to reproduce the results presented in the main text and Supplemental Material.

\bibliography{References}


\clearpage
\onecolumngrid

\setcounter{section}{0}
\setcounter{subsection}{0}
\setcounter{equation}{0}
\setcounter{figure}{0}
\setcounter{table}{0}

\renewcommand{\thesection}{\arabic{section}}
\renewcommand{\thesubsection}{\Alph{subsection}}
\renewcommand{\theequation}{S\arabic{equation}}
\renewcommand{\thefigure}{S\arabic{figure}}
\renewcommand{\thetable}{S\arabic{table}}

\makeatletter
\renewcommand{\p@subsection}{\thesection}
\makeatother

\newcommand{\suppnote}[1]{%
    \refstepcounter{section}%
    \setcounter{subsection}{0}%
    \section*{Supplementary Note \arabic{section}. #1}%
    \addcontentsline{toc}{section}{Supplementary Note \arabic{section}. #1}%
}

\newcommand{\suppsubsection}[1]{%
    \refstepcounter{subsection}%
    \subsection*{\Alph{subsection}. #1}%
    \addcontentsline{toc}{subsection}{\Alph{subsection}. #1}%
}

\section*{Supplementary Material for ``Finite Quantum Engineering Thermodynamics''}

This Supplementary Material contains the derivations, numerical procedures, and validation details needed to support the main text without repeating its discussion. The microscopic density operator remains the exact dynamical state throughout; the thermodynamic state is the smaller measured record used for the declared task.

\suppnote{Relation to the earlier restricted-information framework}
\label{supp:note-lineage}

The present construction continues two earlier restricted-information studies. Reference~\cite{Ahmadi2026Endogenous} starts from a fixed projective record and separates its exact future into the part generated from the present record and a residual carried by microscopic information hidden from that record. In the short-time limit, the record-only contribution is stationary to first order, so unresolved inter-sector currents carry the instantaneous motion of the coarse record.

Reference~\cite{Ahmadi2026FiniteParticle} adds a physical thermodynamic boundary, the retained mean energy, and explicit control. For the finite Bose--Hubbard working medium, the pair \((E,R)\) is sufficient to construct the restricted thermodynamic description used to close a complete cycle without reconstructing the microscopic density operator.

Both works take the accessible record as given. Here the record itself becomes an engineering choice. We ask which experimentally accessible variables should be retained for a specified prediction task, whether unresolved dynamics is better represented by additional instantaneous observables or by finite history, and whether the resulting reduced description can be certified over a declared preparation and operating domain. The new layer is therefore state selection and certification, not a new microscopic dynamics or a new definition of heat and work.

The construction also draws on classical constitutive thermodynamics. Internal variables and history are established alternatives for describing missing dynamics~\cite{ColemanGurtin1967,Day1972,Muschik1990}. In theories of materials with memory, states may be identified by their response to future processes~\cite{Noll1972,FabrizioMorro2005,DeseriFabrizioGolden2006}. Projection and optimal-prediction methods explain how unresolved variables can leave memory in measured dynamics~\cite{Zwanzig1960,Mori1965,Chorin2000,Chorin2002,Rupe2022}. The FQET procedure does not claim to invent these principles. It chooses and certifies finite quantum measurements for a specified task, while retaining the exact Hamiltonian work accounting.

The present results address isolated controlled strokes. They do not construct a general dissipative constitutive theory or derive the Clausius--Duhem restrictions imposed on classical dissipative response functions~\cite{ColemanNoll1963,ColemanGurtin1967}. The certificate here bounds errors in declared future observables. It should not be read as an independent entropy-production inequality. This distinction also limits what is meant by a ``thermodynamic state'': it is a task-sufficient measured record, not a complete state for every admissible future process.

\suppnote{Exact operator-span closure and determinacy}
\label{supp:note-closure}

\suppsubsection{Global determinacy}

Let the retained record be defined by Hermitian observables \(\mathcal R=\{A_1,\ldots,A_m\}\). Including normalization, the information carried linearly by that record is the real operator span
\begin{equation}
\mathcal V_{\mathcal R}
=
\operatorname{span}_{\mathbb R}
\left\{
\mathbb I,A_1,\ldots,A_m
\right\}.
\label{supp:eq:record-span}
\end{equation}

For a Hermitian target \(X\), exact determinacy over the full quantum state space is equivalent to
\begin{equation}
\langle X\rangle
\text{ is determined by the record for every state}
\quad\Longleftrightarrow\quad
X\in\mathcal V_{\mathcal R}.
\label{supp:eq:determinacy}
\end{equation}

The forward implication is immediate. Conversely, if \(X\notin\mathcal V_{\mathcal R}\), let \(X_\perp\neq0\) be its Hilbert--Schmidt component orthogonal to \(\mathcal V_{\mathcal R}\). Because \(\mathbb I\in\mathcal V_{\mathcal R}\), \(X_\perp\) is traceless. For sufficiently small \(\epsilon>0\), the states \(\rho_\pm=\mathbb I/d\pm\epsilon X_\perp\) are positive, have unit trace, agree on every retained observable, and satisfy \(\Tr[X(\rho_+-\rho_-)]>0\). Hence no operator outside the retained span is globally determined.
A restricted preparation class \(\mathfrak P\) can weaken this condition. For a measured record \(\mathbf m\), exact determinacy on the physically allowed preparation fiber requires only
\begin{equation}
\Tr\left[X(\rho-\sigma)\right]=0
\quad
\text{for all }
\rho,\sigma\in
\mathcal A_{\mathcal R,\mathfrak P}(\mathbf m).
\label{supp:eq:restricted-determinacy}
\end{equation}

Thus the operator-span theorem is the exact global statement; preparation information can reduce the remaining freedom further.

\suppsubsection{Work and future-observable closure}

For an isolated Hamiltonian \(H(\boldsymbol\lambda)\), the instantaneous generalized force associated with a controlled parameter \(\lambda_a\) is exactly available from the retained record whenever
\begin{equation}
\frac{\partial H}{\partial\lambda_a}
\in
\mathcal V_{\mathcal R}.
\label{supp:eq:online-work-closure}
\end{equation}

For a prescribed finite protocol \(u\), with propagator \(U_u\), initial Hamiltonian \(H_i\), and final Hamiltonian \(H_f\), the mean work done on the system is
\begin{equation}
W_{\rm on}[\rho,u]
=
\Tr\left[\rho X_W(u)\right],
\quad
X_W(u)
=
U_u^\dagger H_f U_u-H_i.
\label{supp:eq:work-operator}
\end{equation}

Exact prospective work prediction from the initial record is then equivalent to
\begin{equation}
W_{\rm on}[\rho,u]
\text{ is fixed by the initial record for every state}
\quad\Longleftrightarrow\quad
X_W(u)\in\mathcal V_{\mathcal R}.
\label{supp:eq:finite-work-closure}
\end{equation}

The same reasoning applies to any selected output space. If \(\mathcal V_{\rm out}\) is the real span of the required output observables and \(\mathcal V_{\rm in}\) the input record span, exact finite-time prediction is equivalent to
\begin{equation}
U_u^\dagger
\mathcal V_{\rm out}
U_u
\subseteq
\mathcal V_{\rm in}.
\label{supp:eq:prediction-closure}
\end{equation}

For a general process map \(\Phi_u\), one replaces the unitary pullback by \(\Phi_u^\dagger\). By contrast, exact closure for all times under an autonomous Hamiltonian requires the stronger invariance condition
\begin{equation}
i[H,\mathcal V_{\mathcal R}]
\subseteq
\mathcal V_{\mathcal R}.
\label{supp:eq:alltime-closure}
\end{equation}

Generic finite systems need not satisfy this small-space invariance. This is why the present theory is formulated for a declared task and finite horizon rather than for exact all-time closure.

\suppnote{Exact ambiguity and preparation-restricted certificates}
\label{supp:note-ambiguity}

\suppsubsection{Exact global ambiguity}

When exact closure fails, the relevant quantity is the distance of the target from the retained operator space,
\begin{equation}
d_{\mathcal R}(X)
=
\inf_{A\in\mathcal V_{\mathcal R}}
\|X-A\|_\infty .
\label{supp:eq:dR}
\end{equation}

Because the operator norm is dual to the trace norm, this distance has an exact operational meaning:
\begin{equation}
\sup_{\substack{\rho,\sigma\\
\Tr[(\rho-\sigma)A_\mu]=0\ \text{for all }\mu}}
\left|
\Tr[X(\rho-\sigma)]
\right|
=
2d_{\mathcal R}(X).
\label{supp:eq:global-ambiguity}
\end{equation}

For the upper bound, subtract any \(A\in\mathcal V_{\mathcal R}\) and use \(\|\rho-\sigma\|_1\le2\); minimization gives \(2d_{\mathcal R}(X)\). Equality follows from the dual witness for the distance to the subspace: since \(\mathbb I\in\mathcal V_{\mathcal R}\), an optimal witness is traceless, and its positive and negative Jordan parts can be normalized to two density operators with the same retained record. The same duality holds for finite classical probability distributions, with Hermitian operators replaced by bounded functions. Here the additional quantum structure lies in noncommuting target operators and the measurements needed to access them.

For a projective future record, \(P_S=\sum_{r\in S}P_r\) denotes an event projector. The exact global total-variation width is therefore
\begin{equation}
\Delta_{\rm pred}^{\rm glob}(\mathcal R,U)
=
2\max_S
d_{\mathcal R}\left(U^\dagger P_S U\right).
\label{supp:eq:global-TV}
\end{equation}

The corresponding width for finite-protocol work is
\begin{equation}
\Delta W_{\rm glob}(\mathcal R,u)
=
2d_{\mathcal R}\left(X_W(u)\right).
\label{supp:eq:global-work-width}
\end{equation}

\suppsubsection{Preparation-restricted certificates}

Suppose the actual microscopic state lies within trace distance \(\epsilon_{\rm prep}\) of a nominal state \(\rho_\star\) and shares the retained record. Then
\begin{equation}
\left|
\Tr[(\rho-\rho_\star)X]
\right|
\le
2\epsilon_{\rm prep}d_{\mathcal R}(X).
\label{supp:eq:one-sided-tube}
\end{equation}

For two independently admissible states in the same tube, the pairwise width obeys
\begin{equation}
\operatorname{diam}_X
\le
\kappa d_{\mathcal R}(X),
\quad
\kappa
=
\min\left(2,4\epsilon_{\rm prep}\right).
\label{supp:eq:pairwise-tube}
\end{equation}

For the value used in both numerical platforms, \(\epsilon_{\rm prep}=0.01\), one has \(\kappa=0.04\). This is a sufficient upper certificate, not an impossibility theorem: failure of the bound to fall below the task tolerance only means that this bound is inconclusive. Insufficiency requires an admissible lower witness above the tolerance.

\suppsubsection{Positivity-aware refinement}

The generic tube bound does not fully exploit positivity. A tighter state-dependent calculation can instead use the physical compatibility fiber
\begin{equation}
\mathcal C_{\mathcal R,\epsilon}(\rho_0)
=
\left\{
\rho\succeq0:
\Tr\rho=1,\;
\Tr[A_\mu(\rho-\rho_0)]=0\ \text{for all }\mu,\;
D_{\rm tr}(\rho,\rho_0)\le\epsilon
\right\}.
\label{supp:eq:restricted-fiber}
\end{equation}

For a Hermitian target, define its one-sided variation by \(s_{\mathcal R,\epsilon}(X)=\sup_{\rho\in\mathcal C_{\mathcal R,\epsilon}(\rho_0)}\Tr[(\rho-\rho_0)X]\). Its dual representation is
\begin{equation}
s_{\mathcal R,\epsilon}(X)
=
\inf_{\substack{Y\in\mathcal V_{\mathcal R}\\ Z\succeq0}}
\left[
\Tr(\rho_0Z)
+
2\epsilon
\|X-Y+Z\|_\infty
\right].
\label{supp:eq:restricted-dual}
\end{equation}

The exact restricted pairwise diameter is then
\begin{equation}
\operatorname{diam}_{\mathcal C}(X)
=
s_{\mathcal R,\epsilon}(X)
+
s_{\mathcal R,\epsilon}(-X).
\label{supp:eq:restricted-diameter}
\end{equation}

The generic tube bound is used as the common robust certificate in the main numerical comparisons; the positivity-aware form is available when a tighter preparation-specific statement is needed.

\suppnote{Finite-horizon dynamics, currents, and state--history duality}
\label{supp:note-dynamics}

\suppsubsection{Local dynamical layers}

For a projective record with \(p_r(t)=\Tr[\rho(t)P_r]\), the first derivative is
\begin{equation}
\dot p_r(t)
=
\Tr\left[\rho(t)D_r^{(1)}(t)\right],
\quad
D_r^{(1)}(t)
=
i[H(t),P_r].
\label{supp:eq:first-derivative}
\end{equation}

and for a time-dependent Hamiltonian the second derivative is
\begin{equation}
\ddot p_r(t)
=
\Tr\left[\rho(t)D_r^{(2)}(t)\right],
\quad
D_r^{(2)}(t)
=
i[\dot H(t),P_r]
-
[H(t),[H(t),P_r]].
\label{supp:eq:second-derivative-general}
\end{equation}

Thus a local finite-horizon expansion reads
\begin{equation}
p_r(t_c+\tau)
=
p_r(t_c)
+
\tau\dot p_r(t_c)
+
\frac{\tau^2}{2}\ddot p_r(t_c)
+
\mathcal O(\tau^3).
\label{supp:eq:future-taylor}
\end{equation}

The first layer measures the current of probability between record sectors and is the local form of the hidden-current structure identified in Ref.~\cite{Ahmadi2026Endogenous}. Higher derivatives continue the same hierarchy. In both driven models used here, \(\dot H\) is proportional to the operator resolved by the record projectors, so \([\dot H,P_r]=0\) and the explicit \(i[\dot H,P_r]\) term vanishes.

\suppsubsection{History as dynamical information}

A record measured at an earlier time can be represented at the decision time by a back-transported operator,
\begin{equation}
p_r(t_c-s)
=
\Tr\left[
\rho(t_c)
\,P_r^{(-s)}(t_c)
\right],
\quad
P_r^{(-s)}(t_c)
=
U(t_c,t_c-s)P_rU^\dagger(t_c,t_c-s).
\label{supp:eq:past-as-current-observable}
\end{equation}

For a short lag,
\begin{equation}
p_r(t_c-s)
=
p_r(t_c)
-
s\dot p_r(t_c)
+
\frac{s^2}{2}\ddot p_r(t_c)
+
\mathcal O(s^3).
\label{supp:eq:history-taylor}
\end{equation}

so nearby history can recover the local slope and curvature through finite differences. This can be expensive because nearly equal estimated probabilities must be subtracted. At longer lags, the back-transported projectors need not lie near the span of the retained first and second derivative layers and can carry useful finite-time information that those layers miss. In an analytic finite-dimensional model, sufficiently high derivatives can still reconstruct the finite-lag operator; the distinction is from a truncated local derivative space, not from the full hierarchy.

This is the state--history choice used below: missing dynamics may be represented either by enlarging the instantaneous state or by retaining a finite history. Similar choices occur in classical constitutive and optimal-prediction theories~\cite{ColemanGurtin1967,Chorin2002,Rupe2022}. Here the historical values are estimated on independently prepared copies; they are not sequential projective measurements on one undisturbed state. The representations can have different predictive power, measurement cost, and conditioning.

\suppnote{Constitutive accuracy, metrology, and task-aware optimization}
\label{supp:note-metrology}

\suppsubsection{One finite-confidence task budget}

Let \(X_k\) be a required pulled-back target and let its constitutive approximation be built from the retained observables. Under the ideal repeated-projective benchmark, a setting \(s\) is measured on \(N_s\) independently prepared copies. Hoeffding bounds are applied simultaneously to the complete family of required estimates, so each setting contributes a term proportional to \(N_s^{-1/2}\). For a family of \(M\) simultaneous estimates with total failure probability \(\alpha\), we collect the estimator range \(\Delta Y_{ks}\) and confidence factor into
\begin{equation}
\frac{b_{ks}(c_k)}{\sqrt{N_s}},
\quad
b_{ks}(c_k)
=
\sqrt{2\ln(2M/\alpha)}
\,\Delta Y_{ks}(c_k).
\label{supp:eq:setting-burden}
\end{equation}

The finite-confidence certificate used in the optimization is
\begin{equation}
\Gamma_k
=
\delta_{{\rm lift},k}
+
\kappa
\left\|
X_k-a_{0k}\mathbb I-\sum_\mu c_{k\mu}A_\mu
\right\|_\infty
+
\sum_s
\frac{b_{ks}(c_k)}{\sqrt{N_s}}.
\label{supp:eq:taskaware-certificate}
\end{equation}

Here \(\delta_{{\rm lift},k}\) is a deterministic allowance for extending a calibrated discrete calculation to the requested continuous time or parameter point; it is zero when no such interpolation or lifting is needed. The second term is the preparation-restricted constitutive ambiguity, and the last term is the finite-confidence measurement uncertainty. All three consume the same task tolerance.

\suppsubsection{Optimal repetitions for a fixed constitutive representation}

For one active constraint, define the remaining margin \(m\) after the intrinsic and lifting contributions have been paid. Minimizing \(\sum_sN_s\) subject to \(\sum_s b_s/\sqrt{N_s}\le m\) gives
\begin{equation}
N_s^\star
=
\frac{
\left(\sum_j b_j^{2/3}\right)^2
b_s^{2/3}
}{m^2},
\quad
N_{\rm tot}^\star
=
\frac{
\left(\sum_s b_s^{2/3}\right)^3
}{m^2}.
\label{supp:eq:shot-allocation}
\end{equation}

Equivalently,
\begin{equation}
N_{\rm tot}^\star
=
\frac{B}{m^2}.
\label{supp:eq:burden-margin-factorization}
\end{equation}

where \(B=(\sum_s b_s^{2/3})^3\). The factorization makes the tradeoff transparent: a representation can be expensive because it leaves too little margin \(m\), because its coefficients create a large metrological burden \(B\), or both.

\suppsubsection{Joint state and metrology design}

The actual task contains many future times and events sharing the same measurement settings. We therefore optimize the constitutive coefficients, history lags when present, and common shot allocation together,
\begin{equation}
\min_{\{c_k\},\{N_s\},\{s_j\}}
\sum_sN_s
\quad
\text{subject to}
\quad
\Gamma_k\le\epsilon_{\rm task}
\quad
\text{for every required constraint }k.
\label{supp:eq:taskaware-joint-opt}
\end{equation}

For fixed lags and fixed metrological weights, the coefficient subproblems have the convex spectral structure used in the numerical implementation. The complete outer search also contains lag placement, support changes, and shared-shot reallocation and is therefore treated as a screened and refined engineering optimization rather than a theorem-level global optimization.

The repeated-projective model is a conservative common comparison benchmark, not a detector-level prediction. A quoted repetition advantage depends on the stipulated observables, shot model, preparation set, and forecasting task; it is not a universal preference for history or derivatives. A specific experiment can replace it with its own readout noise, backaction, latency, preparation rate, and setting cost without changing the operator-level framework.

\subsection{A worked qubit example}
\label{sec:qubit_example}

A single qubit makes the different error sources clear before we turn to many-body systems. The example also shows why history and a direct current can contain the same information while carrying different measurement costs.

Consider
\begin{equation}
H(\lambda)
=
\frac{\omega}{2}\sigma_x
+
\frac{\lambda}{2}\sigma_z ,
\end{equation}
where $\lambda$ is externally controlled.  Write the microscopic state as
\begin{equation}
\rho
=
\frac{1}{2}
\left(
I+x\sigma_x+y\sigma_y+z\sigma_z
\right).
\end{equation}
Suppose the coarse thermodynamic record is the projective measurement of $\sigma_z$,
\begin{equation}
P_{\pm}
=
\frac{I\pm\sigma_z}{2},
\quad
p_{\pm}
=
\Tr(\rho P_{\pm})
=
\frac{1\pm z}{2}.
\end{equation}
The record already closes the work associated with the controlled parameter $\lambda$, because
\begin{equation}
\frac{\partial H}{\partial\lambda}
=
\frac{\sigma_z}{2}.
\end{equation}
Hence an isolated infinitesimal control step satisfies
\begin{equation}
\delta W_{\rm on}
=
\frac{z}{2}\,d\lambda,
\quad
\delta Q=0,
\end{equation}
without reconstructing the microscopic state.

Now consider a forecasting task.  At a decision time $t_c$, set $\lambda=0$, so that the subsequent Hamiltonian is
\begin{equation}
H_0
=
\frac{\omega}{2}\sigma_x .
\end{equation}
If the mean energy is also retained, the present information determines $x$ and $z$, and its operator span is
\begin{equation}
V_{E,R}
=
\operatorname{span}_{\mathbb R}
\{I,\sigma_x,\sigma_z\}.
\end{equation}
The only Pauli direction missing from this span is $\sigma_y$.
The future probability of the $+$ outcome after a time $\tau$ can be written at the decision time as
\begin{equation}
p_+(t_c+\tau)
=
\Tr(\rho X_\tau),
\quad
X_\tau
=
U^\dagger(\tau)P_+U(\tau),
\end{equation}
where $U(\tau)=e^{-iH_0\tau}$.  Direct evaluation gives
\begin{equation}
X_\tau
=
\frac{1}{2}
\left[
I
+
\cos(\omega\tau)\sigma_z
+
\sin(\omega\tau)\sigma_y
\right].
\end{equation}
Equivalently,
\begin{equation}
z(t_c+\tau)
=
z(t_c)\cos(\omega\tau)
+
y(t_c)\sin(\omega\tau).
\end{equation}
The present state $(E,R)$ therefore does not determine the future record whenever $\sin(\omega\tau)\neq0$, because it does not determine $y$.
The part of the target already contained in the retained span is
\begin{equation}
A_\tau
=
\frac{1}{2}
\left[
I+\cos(\omega\tau)\sigma_z
\right],
\end{equation}
so that the unresolved operator direction is
\begin{equation}
X_\tau-A_\tau
=
\frac{\sin(\omega\tau)}{2}\sigma_y,
\quad
\|X_\tau-A_\tau\|_\infty
=
\frac{|\sin(\omega\tau)|}{2}.
\end{equation}
This is the microscopic information absent from the present thermodynamic state.
The preparation factor $\kappa$ has a direct meaning in this example.  Let the nominal state be $\rho_\star=I/2$ and suppose every admissible preparation obeys
\begin{equation}
D_{\rm tr}(\rho,\rho_\star)
\le
\epsilon_{\rm prep}.
\end{equation}
Two admissible states can be chosen as
\begin{equation}
\rho_\pm
=
\frac{I}{2}
\pm
\epsilon_{\rm prep}\sigma_y.
\end{equation}
They have the same present energy and record, but their trace-norm separation is
\begin{equation}
\|\rho_+-\rho_-\|_1
=
4\epsilon_{\rm prep}.
\end{equation}
Since two arbitrary density operators can never have trace-norm distance larger than $2$, the generic pairwise preparation factor is
\begin{equation}
\kappa
=
\min(2,4\epsilon_{\rm prep}).
\end{equation}
For the pair above,
\begin{equation}
\left|
\Tr\!\left[
X_\tau(\rho_+-\rho_-)
\right]
\right|
=
2\epsilon_{\rm prep}
|\sin(\omega\tau)|
=
\kappa
\|X_\tau-A_\tau\|_\infty
\end{equation}
when $4\epsilon_{\rm prep}\le2$.  Thus $\kappa$ controls how far two preparation-compatible microscopic states may separate, while $\|X_\tau-A_\tau\|_\infty$ measures how strongly the target responds to the unresolved direction.
The missing $\sigma_y$ information can be supplied in two ways.  The instantaneous current operator is
\begin{equation}
D_+^{(1)}
=
i[H_0,P_+]
=
\frac{\omega}{2}\sigma_y,
\end{equation}
with
\begin{equation}
J
=
\left\langle D_+^{(1)}\right\rangle
=
\frac{\omega}{2}y.
\end{equation}
Retaining $J$ therefore closes the future probability exactly:
\begin{equation}
p_+(t_c+\tau)
=
\frac{1}{2}
\left[
1
+
z(t_c)\cos(\omega\tau)
+
\frac{2J(t_c)}{\omega}\sin(\omega\tau)
\right].
\end{equation}

The same information can instead be supplied by one past record.  Let $\ell>0$ denote a history lag.  Exact evolution gives
\begin{equation}
z(t_c-\ell)
=
z(t_c)\cos(\omega\ell)
-
y(t_c)\sin(\omega\ell),
\end{equation}
and hence, when $\sin(\omega\ell)\neq0$,
\begin{equation}
y(t_c)
=
\frac{
z(t_c)\cos(\omega\ell)-z(t_c-\ell)
}{
\sin(\omega\ell)
}.
\end{equation}
The historical representation therefore also closes the forecasting task exactly.
Exact closure, however, does not imply equal measurement cost.  Substituting the historical expression into the future probability gives
\begin{equation}
p_+(t_c+\tau)
=
\frac{1}{2}
+
c_0(\tau,\ell)z(t_c)
+
c_\ell(\tau,\ell)z(t_c-\ell),
\end{equation}
where
\begin{equation}
c_\ell(\tau,\ell)
=
-
\frac{1}{2}
\frac{\sin(\omega\tau)}
{\sin(\omega\ell)}.
\end{equation}
A projective $\sigma_z$ measurement has single-shot outcomes $\pm1$.  The single-shot contribution from the historical setting therefore has range
\begin{equation}
\Delta Y_\ell
=
2|c_\ell|
=
\frac{|\sin(\omega\tau)|}
{|\sin(\omega\ell)|}.
\end{equation}
Using the definition in Eq.~\eqref{supp:eq:setting-burden},
\begin{equation}
b_\ell
=
\sqrt{2\ln(2M/\alpha)}
\,
\frac{|\sin(\omega\tau)|}
{|\sin(\omega\ell)|}.
\end{equation}
For a short lag,
\begin{equation}
b_\ell
\sim
\sqrt{2\ln(2M/\alpha)}
\,
\frac{|\sin(\omega\tau)|}
{\omega\ell},
\quad
\omega\ell\ll1.
\end{equation}
Thus history determines the future probability exactly when the denominator is nonzero, but its statistical burden diverges as $\ell\rightarrow0$. Physically, the current is recovered by subtracting two nearly equal measured averages and dividing by their small time separation. Repeating each measurement more accurately can compensate for this amplification, but doing so raises the experimental cost.

By contrast, direct measurement of $\sigma_y$ has single-shot outcomes $\pm1$.  Its coefficient in the future probability is $\sin(\omega\tau)/2$, giving
\begin{equation}
\Delta Y_y
=
|\sin(\omega\tau)|,
\quad
b_y
=
\sqrt{2\ln(2M/\alpha)}
|\sin(\omega\tau)|.
\end{equation}
The direct current representation therefore avoids the $1/\sin(\omega\ell)$ amplification present in short-lag history.
This example makes the two non-lifting terms in Eq.~(S24) transparent.  The quantity
\begin{equation}
\kappa\|X-A\|_\infty
\end{equation}
measures uncertainty associated with microscopic information that was \emph{not} retained, and cannot be reduced by repeating the retained measurements.  In contrast,
\begin{equation}
\frac{b_s}{\sqrt{N_s}}
\end{equation}
measures statistical uncertainty in information that \emph{was} retained, and decreases with additional repetitions.  In the present analytic example no interpolation is needed, so $\delta_{\rm lift}=0$.

The qubit also shows why a richer candidate state need not acquire a new active coordinate.  Its next local dynamical layer is
\begin{equation}
D_+^{(2)}
=
-[H_0,[H_0,P_+]]
=
-\frac{\omega^2}{2}\sigma_z,
\end{equation}
which already lies in the present record span.  Adding it therefore supplies no new operator direction.  In the many-body systems studied below, higher dynamical layers and additional history nodes need not be redundant.  We therefore next distinguish the coordinates that are merely available from those that become genuinely useful for the declared task.

\suppnote{Nested supports and task-aware state selection}
\label{supp:note-support}

The two representation families are nested,
\begin{equation}
(E,R,\mathcal J)
\subset
(E,R,\mathcal J,K^{(2)}),
\quad
H_2
\subset
H_3.
\label{supp:eq:nested-supports}
\end{equation}

A richer candidate can always reproduce the smaller one by assigning zero constitutive weight to its added coordinates. Hence, after a correct optimization,
\begin{equation}
C^\star(\mathcal S_+)
\le
C^\star(\mathcal S_-).
\label{supp:eq:nested-cost}
\end{equation}

A numerical calculation can nevertheless miss that relation if the two supports are optimized from unrelated starting points. We therefore cross-seed in both directions: the smaller solution is embedded into the larger family with zero new coefficients, while the larger solution is projected back and reoptimized.

An added layer is called \emph{active} only if it acquires a finite measurement role and lowers the best cross-seeded cost. This distinguishes candidate support from active support and prevents optimizer-basin changes from being mistaken for physical transitions. Equal coordinate count is used only as a fairness check; it is not a claim of equal laboratory cost.
Family selection and layer activation are different questions. We compare the lower envelopes
\begin{equation}
C_{\rm hist}
=
\min\left(C_{H_2},C_{H_3}\right),
\quad
C_{\rm inst}
=
\min\left(C_{\mathcal J},C_{\mathcal J K}\right).
\label{supp:eq:family-envelopes}
\end{equation}

so a second history node can become useful even when the history family is not the lower-shot family, and similarly for \(K^{(2)}\). Mandatory thermodynamic coordinates, such as energy when required for work accounting, are retained even if the forecasting subproblem assigns them little extra weight.

Finally, repetition count is only one resource. Setting count, calibration effort, latency, disturbance, and hardware requirements remain separate unless a concrete experiment supplies a justified scalar cost. When resources disagree, the output is a Pareto tradeoff rather than an artificial universal winner.

\suppnote{Bose--Hubbard working medium and microscopic numerical construction}
\label{supp:note-bh-microscopic}

\suppsubsection{Model, record, and protocol}

We use \(Q=4\) bosons on an open chain of \(L=4\) sites, restricted to the fixed-particle Fock basis \(\ket{n_1,n_2,n_3,n_4}\) with \(\sum_jn_j=4\). Its dimension is \(\binom{Q+L-1}{Q}=35\). The Hamiltonian is
\begin{equation}
H(J,U,\Lambda)
=
-JT
+
\frac{U}{2}D
+
\Lambda N_R,
\label{supp:eq:bh-hamiltonian}
\end{equation}

with \(T=\sum_j(a_j^\dagger a_{j+1}+a_{j+1}^\dagger a_j)\), \(D=\sum_j n_j(n_j-1)\), and \(N_R=n_3+n_4\). Let \(P_r\) project onto the sector with \(r\) particles in the right half. The retained spatial record is
\begin{equation}
R(t)
=
\{p_0(t),\ldots,p_4(t)\},
\quad
p_r(t)
=
\Tr[\rho(t)P_r].
\label{supp:eq:bh-record}
\end{equation}

and it closes the generalized force of the controlled confinement because
\begin{equation}
\frac{\partial H}{\partial\Lambda}
=
N_R
=
\sum_r rP_r.
\label{supp:eq:bh-lambda-force}
\end{equation}

The nominal initial state is canonical at \(\kb T_h=5J\), with \(U_0/J=1.3/3\) and \(\Lambda_{0,0}=20J\):
\begin{equation}
\rho_0
=
\frac{
e^{-H(J,U,\Lambda_0)/(\kb T_h)}
}{
\Tr\left[
e^{-H(J,U,\Lambda_0)/(\kb T_h)}
\right]
}.
\label{supp:eq:bh-canonical}
\end{equation}

After preparation the system is isolated and the confinement is ramped linearly,
\begin{equation}
\Lambda(t)
=
\Lambda_0
\left(
1-\frac{t}{\tau_{BC}}
\right),
\quad
0\le t\le\tau_{BC}.
\label{supp:eq:bh-ramp}
\end{equation}

with nominal \(J\tau_{BC,0}=20\). The isolated energy change is checked against the generalized-force work,
\begin{equation}
E(t_2)-E(t_1)
=
\int_{t_1}^{t_2}
\dot\Lambda(t)
\langle N_R\rangle_t\,dt
=
W_{\rm on}(t_1\rightarrow t_2).
\label{supp:eq:bh-energy-balance}
\end{equation}

A uniform forecast bound on the spatial record also controls a physical work prediction. If two admissible future records differ by at most \(\epsilon_{\rm TV}\) in total variation at every time in a subinterval of the ramp, then \(N_R\) takes values between zero and four and their mean right-half occupations differ by at most \(4\epsilon_{\rm TV}\). Using Eq.~\eqref{supp:eq:bh-energy-balance}, their predicted works therefore satisfy
\begin{equation}
|W_{\rm on}^{(1)}-W_{\rm on}^{(2)}|
\le
4\epsilon_{\rm TV}
\int_{t_1}^{t_2}
|\dot\Lambda(t)|\,dt.
\label{supp:eq:bh-work-from-record}
\end{equation}
This is a consequence of the controlled-drive energy balance, not a new constitutive definition of work. It applies when the total-variation bound holds uniformly on the same interval and concerns the difference between the two admissible records.

The prediction is made at
\begin{equation}
Jt_c
=
0.2,
\quad
0\le J\tau\le0.2.
\label{supp:eq:bh-decision-horizon}
\end{equation}

with preparation radius \(\epsilon_{\rm prep}=0.01\), target pairwise total-variation width \(1\%\), and the common \(99\%\) family-wise confidence benchmark.

\suppsubsection{Prediction operators and microscopic numerics}

For an event \(S\subset\{0,1,2,3,4\}\), \(P_S=\sum_{r\in S}P_r\). Its future probability is represented at the decision time by
\begin{equation}
p_S(t_c+\tau)
=
\Tr\left[
\rho(t_c)X_S(\tau)
\right],
\quad
X_S(\tau)
=
U_\tau^\dagger P_SU_\tau.
\label{supp:eq:bh-target-operator}
\end{equation}

Because complementary events give the same absolute probability difference, only \(15\) independent binary events are required to control the total-variation distance of the five-outcome record.

The instantaneous candidate family uses
\begin{equation}
D_r^{(1)}(t_c)
=
i[H(t_c),P_r],
\quad
D_r^{(2)}(t_c)
=
-[H(t_c),[H(t_c),P_r]].
\label{supp:eq:bh-derivative-operators}
\end{equation}

while a history value at lag \(s\) is represented by
\begin{equation}
P_r^{(-s)}(t_c)
=
U(t_c,t_c-s)
P_r
U^\dagger(t_c,t_c-s).
\label{supp:eq:bh-history-operator}
\end{equation}

The calibration propagator is generated with adaptive DOP853 using relative tolerance \(2\times10^{-9}\) and absolute tolerance \(2\times10^{-11}\). The Fock basis, projector completeness, Hermiticity, state trace and positivity, and work--energy equality are checked independently. Reconstructing the nominal endpoint reproduces the earlier continuous-ramp calculation at the \(10^{-9}\)--\(10^{-8}\) level for the displayed observables.
Historical probabilities are interpreted as stored ensemble estimates. Under the repeated-projective benchmark, different times and incompatible observables are estimated from independently prepared copies; no undisturbed projective history measurement on a single realization is assumed.

\suppnote{Task-aware selection of the Bose--Hubbard thermodynamic state}
\label{supp:note-bh-selection}

\suppsubsection{Candidate representations and final comparison}

The baseline thermodynamic information is \((E,R)\). To close the future record we compare two equal-coordinate nested families: the instantaneous states \((E,R,\mathcal J)\) and \((E,R,\mathcal J,K^{(2)})\), and the history states \(H_2\) and \(H_3\). The constitutive coefficients, lags, and shared metrology are optimized jointly, with the cross-seeding rule of Note~\ref{supp:note-support}.

The final nominal comparison is given in Table~\ref{supp:tab:bh-taskaware-comparison}.

\begin{table}[t]
\centering
\small
\begin{tabular}{ccccc}
\hline\hline
\(J\tau\)
&
Best history
&
History shots
&
Best instantaneous state
&
Instantaneous shots
\\
\hline
0.050
&
\(H_2(0.20)\)
&
\(1.66602\times10^6\)
&
\((E,R,\mathcal J)\)
&
\(3.20580\times10^6\)
\\
0.075
&
\(H_2(0.20)\)
&
\(1.96181\times10^6\)
&
\((E,R,\mathcal J,K^{(2)})\)
&
\(6.81726\times10^6\)
\\
0.100
&
\(H_2(0.20)\)
&
\(1.96331\times10^6\)
&
\((E,R,\mathcal J,K^{(2)})\)
&
\(1.21336\times10^7\)
\\
0.150
&
\(H_2(0.20)\)
&
\(2.19456\times10^6\)
&
\((E,R,\mathcal J,K^{(2)})\)
&
\(1.91534\times10^7\)
\\
0.200
&
\(H_3(0.10,0.20)\)
&
\(2.49002\times10^6\)
&
\((E,R,\mathcal J,K^{(2)})\)
&
\(2.12532\times10^7\)
\\
\hline\hline
\end{tabular}
\caption{
Final task-aware Bose--Hubbard representation comparison at the nominal
operating point. The constitutive coefficients and shared repeated-projective
metrology are optimized together, and nested candidate supports are
cross-seeded before the active support is assigned. The numbers are
model-, task-, and metrology-specific engineering results rather than
universal measurement costs.
}
\label{supp:tab:bh-taskaware-comparison}
\end{table}

History is lower-shot throughout the tested range. At \(J\tau=0.20\), the best instantaneous architecture requires approximately \(8.54\) times as many repetitions as the best history architecture.

\suppsubsection{Support activation and selected state}

The direct second-derivative layer becomes active at
\begin{equation}
0.051
<
J\tau_{K^{(2)}}
<
0.052.
\label{supp:eq:bh-k2-activation}
\end{equation}

whereas the second history node activates much later,
\begin{equation}
0.150
<
J\tau_{H_3}
<
0.1525.
\label{supp:eq:bh-h3-activation}
\end{equation}

This ordering does not make the instantaneous family preferable: activation inside a family and selection between families are distinct decisions.

For one fixed online architecture that covers the complete design horizon, we therefore retain
\begin{equation}
\mathcal R_{\rm BH}(t_c)
=
\left\{
E(t_c),
R(t_c),
R(t_c-0.1/J),
R(t_c-0.2/J)
\right\}.
\label{supp:eq:bh-selected-state}
\end{equation}

The five-outcome record contributes four independent probabilities, so \(H_2\) and \((E,R,\mathcal J)\) each contain nine independent scalars, while \(H_3\) and \((E,R,\mathcal J,K^{(2)})\) each contain thirteen. The history advantage is therefore not created by giving that family more coordinates. It is a model-, task-, and metrology-specific engineering result, not a universal preference for memory.

\suppnote{Continuous Bose--Hubbard constitutive data sheet}
\label{supp:note-bh-datasheet}

\suppsubsection{Frozen state and parameterized response}

The state in Eq.~\eqref{supp:eq:bh-selected-state} is fixed before the device-parameter scan. The declared operating box is
\begin{equation}
\frac{U}{U_0},
\quad
\frac{\Lambda_0}{\Lambda_{0,0}},
\quad
\frac{\tau_{BC}}{\tau_{BC,0}}
\in
[0.9,1.1].
\label{supp:eq:bh-parameter-box}
\end{equation}

For each event and future time, the constitutive prediction is linear in the measured thermodynamic coordinates, with coefficients that depend on the device parameters. Only the eight corners of the three-dimensional box are stored. Between them the coefficients are interpolated trilinearly,
\begin{equation}
c
\left(
\boldsymbol{\theta}
\right)
=
\sum_{i,j,k\in\{0,1\}}
w_i(x_U)
w_j(x_{\Lambda})
w_k(x_{\tau})
c_{ijk},
\label{supp:eq:bh-trilinear}
\end{equation}

where the normalized coordinates run from zero to one along each parameter direction.

\suppsubsection{Continuous-domain validation}

Corner agreement alone does not control interpolation between corners. We therefore separate response calibration from validation. A \(5\times5\times5=125\) lattice, supplemented by deterministic interior checks, is used only to bound parameter sensitivity. The earlier coarser audit was rejected because random interior sensitivities exceeded its maxima by factors about \(1.72\), \(1.77\), and \(2.12\) in the three directions. On the dense grid the additional interior excesses fall to approximately \(0.15\%\), \(0.97\%\), and \(4.79\%\), after which a conservative \(20\%\) inflation is applied.

For a nearby validated anchor \(\boldsymbol\theta_i\), the continuous parameter lift uses the sensitivity envelope in the standard Lipschitz form
\begin{equation}
g
\left(
\boldsymbol{\theta}
\right)
\le
g
\left(
\boldsymbol{\theta}_i
\right)
+
\sum_a
M_a^{\rm env}
\left|
\theta_a-\theta_{i,a}
\right|.
\label{supp:eq:bh-parameter-lift}
\end{equation}

Combining this parameter lift with the continuous-time lift and preparation-tube ambiguity gives the worst continuous intrinsic certificate
\begin{equation}
\Gamma_{\rm int}^{\rm cont}
\le
0.7304\%.
\label{supp:eq:bh-continuous-intrinsic}
\end{equation}

The complete \(99\%\) repeated-projective benchmark then requires
\begin{equation}
N_{\rm shots}
\simeq
3.786\times10^7.
\label{supp:eq:bh-datasheet-shots}
\end{equation}

The response-table count, offline validation density, and online metrology are different resources. Table~\ref{supp:tab:bh-datasheet-comparison} shows how their costs change when the calibration and validation grids are varied.

\begin{table}[t]
\centering
\small
\begin{tabular}{lcccc}
\hline\hline
Response
&
Tables
&
Validation
&
Worst intrinsic
&
Shared shots
\\
\hline
Frozen nominal coefficients
&
1
&
153
&
\(0.9801\%\)
&
\(2.791\times10^9\)
\\
Eight-corner trilinear
&
8
&
27
&
\(0.8861\%\)
&
\(1.886\times10^8\)
\\
\(3^3\) piecewise trilinear
&
27
&
27
&
\(0.8224\%\)
&
\(7.762\times10^7\)
\\
Eight-corner trilinear
&
8
&
125
&
\(0.7304\%\)
&
\(3.786\times10^7\)
\\
\(3^3\) piecewise trilinear
&
27
&
125
&
\(0.7008\%\)
&
\(3.293\times10^7\)
\\
\hline\hline
\end{tabular}
\caption{
Comparison of continuous-domain Bose--Hubbard constitutive constructions.
``Tables'' counts response-calibration tables stored by the constitutive
model, whereas ``Validation'' counts offline parameter anchors used to audit
the continuous-domain certificate. The shared repetition count refers to the
common idealized metrology benchmark. These three resources should not be
identified with one another.
}
\label{supp:tab:bh-datasheet-comparison}
\end{table}

The adopted construction is the eight-corner trilinear response with the \(125\)-point validation audit. The \(0.7304\%\) statement is a continuous-domain engineering certificate under the declared numerical sensitivity envelope, not a formal interval-arithmetic proof of the exact global supremum and not a proof that no alternative architecture could be cheaper.

\suppnote{Blind validation of the frozen Bose--Hubbard data sheet}
\label{supp:note-bh-blind}

Before the holdout test, the selected state, history lags, eight corner tables, interpolation rule, and sensitivity envelope are frozen. No coefficient, lag, response node, or error envelope is changed after the holdout is inspected.

Solver B reconstructs the fixed-particle sector independently and uses a fourth-order two-node Gauss--Magnus propagator with \(J\Delta t=0.0025\), rather than the DOP853 calibration solver. The holdout contains \(64\) scrambled Sobol points plus \(8\) near-corner points. At each of the \(72\) parameter points we test the nominal canonical preparation and three independent positive perturbations near trace distance \(0.0098\), at \(81\) future times and all \(15\) binary events.

The blind-validation results are collected in Table~\ref{supp:tab:bh-blind-summary}.

\begin{table}[t]
\centering
\small
\begin{tabular}{lc}
\hline\hline
Blind-validation quantity & Result \\
\hline
Unseen parameter points & \(72\) \\
Preparations per point & \(4\) \\
Future times per preparation & \(81\) \\
Independent binary events & \(15\) \\
Total blind event predictions & \(349{,}920\) \\
Maximum canonical error & \(0.1197\%\) \\
Maximum tube-state error & \(0.1278\%\) \\
Overall maximum error & \(0.1278\%\) \\
Worst-event-at-each-time 99th percentile & \(0.1148\%\) \\
Maximum unitarity defect & \(1.517\times10^{-14}\) \\
Maximum step-halving event difference & \(2.002\times10^{-13}\) \\
\hline\hline
\end{tabular}
\caption{
Summary of the independent Bose--Hubbard holdout campaign. The prediction
errors are numerical/model-validation errors of the frozen constitutive data
sheet. The final two rows quantify the numerical convergence of Solver B and
are many orders of magnitude below the constitutive prediction errors.
}
\label{supp:tab:bh-blind-summary}
\end{table}

The worst holdout occurs near \((U/U_0,\Lambda_0/\Lambda_{0,0},\tau_{BC}/\tau_{BC,0})=(1.095,0.905,1.095)\) at \(J\tau=0.150\). Across the holdout, the pointwise maximum error is strongly anticorrelated with the confinement scale (Pearson \(-0.9818\), Spearman \(-0.9802\)); the correlations with interaction and duration are weak.

The observed \(0.1278\%\) maximum and the \(0.7304\%\) continuous certificate are different objects. The first is a finite unseen-sample model error; the second is a sufficient continuous-domain upper certificate that also includes preparation and interpolation allowances. The holdout is therefore an independent numerical/model validation, not an experimental validation and not a replacement for the continuous-domain certificate.

\suppnote{Transverse-field Ising working medium and task-aware representation map}
\label{supp:note-ising}

\suppsubsection{Model, record, and candidate states}

For an independent test of representation selection, we use an open chain of four spins with
\begin{equation}
H(t)
=
-JC_{zz}
-gX_{\rm tot}
+\Lambda(t)N_{\uparrow},
\label{supp:eq:ising-H}
\end{equation}

where \(C_{zz}=\sum_{j=1}^{L-1}Z_jZ_{j+1}\), \(X_{\rm tot}=\sum_jX_j\), and \(N_\uparrow=\sum_j(\mathbb I+Z_j)/2\). The initial state is canonical at \(k_{\rm B}T=2J\); the isolated drive ramps \(\Lambda\) linearly from \(+2J\) to \(-2J\) over \(J\tau_{\rm drv}=4\). The transverse field \(g\) is varied between operating points.

The coarse record is
\begin{equation}
R(t)
=
\left\{
p_0(t),\ldots,p_4(t)
\right\},
\quad
p_r(t)
=
\Tr[\rho(t)P_r].
\label{supp:eq:ising-record}
\end{equation}

with \(P_r\) projecting onto the sector with \(r\) up spins. Since \(N_\uparrow=\sum_r rP_r\), the record closes the work associated with the \(\Lambda\) drive. Adding \(E\) and \(C_{zz}\) also closes the remaining static generalized forces for \(g\neq0\), but gives almost no improvement to the short-time future-record prediction. Static work closure and forecasting closure are therefore distinct tasks.

The instantaneous family is
\begin{equation}
\mathcal R_{\rm inst}^{(1)}
=
\left\{
E,R,\mathcal J
\right\},
\quad
\mathcal R_{\rm inst}^{(2)}
=
\left\{
E,R,\mathcal J,K^{(2)}
\right\}.
\label{supp:eq:ising-instantaneous-family}
\end{equation}

and the history family is
\begin{equation}
H_2(s_1)
=
\left\{
E(t_c),
R(t_c),
R(t_c-s_1)
\right\},
\label{supp:eq:ising-H2}
\end{equation}
\begin{equation}
H_3(s_1,s_2)
=
\left\{
E(t_c),
R(t_c),
R(t_c-s_1),
R(t_c-s_2)
\right\},
\quad
0<s_1<s_2\le t_c.
\label{supp:eq:ising-H3}
\end{equation}

The smaller states contain nine independent scalar coordinates and the larger states thirteen. History lags are searched independently of the Bose--Hubbard values.

\suppsubsection{Task and representation boundaries}

The decision time is \(Jt_c=1\), leaving one unit of earlier trajectory available to the history search. The common preparation and prediction targets are
\begin{equation}
\epsilon_{\rm prep}
=
0.01,
\quad
\epsilon_{\rm TV}
=
0.01,
\label{supp:eq:ising-task-tolerances}
\end{equation}

with \(99\%\) family-wise confidence. The future grid uses \(J\Delta t=0.01\). The event-operator Lipschitz bound \(L_{\max}=6J\), together with \(\kappa=0.04\), lifts a half-cell to at most \(0.0012\) in pairwise total variation, so the map controls the complete forecast interval rather than only its endpoint.

Three boundaries are distinguished: \(\tau_{\rm family}\), where the lower-shot family changes while both use their first dynamical layer; \(\tau_{H_3}\), where the second history node activates; and \(\tau_{K^{(2)}}\), where the second direct derivative layer activates. At every tested field,
\begin{equation}
\tau_{\rm family}
<
\tau_{H_3}
<
\tau_{K^{(2)}}.
\label{supp:eq:ising-boundary-ordering}
\end{equation}

The numerical values for the three boundaries are collected in Table~\ref{supp:tab:ising-final-map}.

\begin{table}[t]
\centering
\small
\begin{tabular}{cccc}
\hline\hline
\(g/J\)
&
\(J\tau_{\rm family}\)
&
\(J\tau_{H_3}\)
&
\(J\tau_{K^{(2)}}\)
\\
\hline
0.75
&
0.05367
&
0.09705
&
0.135--0.140
\\
1.00
&
0.04220
&
0.06859
&
0.105--0.1075
\\
1.25
&
0.03461
&
0.05550
&
0.085--0.0875
\\
1.50
&
0.029207
&
0.041083
&
0.07150--0.07175
\\
1.75
&
0.02489
&
0.03552
&
0.060--0.0625
\\
2.00
&
0.02148
&
0.02870
&
0.0525--0.05375
\\
\hline\hline
\end{tabular}
\caption{
Final task-aware Ising representation map. The family boundary is the
shared-repetition equality between \(H_2\) and \((E,R,\mathcal J)\).
The \(H_3\) column marks activation of the second history layer, while the
\(K^{(2)}\) column marks activation of the direct second-derivative layer.
The latter is reported as a validated sign-changing bracket where that is the
numerical resolution supported by the calculation.
}
\label{supp:tab:ising-final-map}
\end{table}

At \(g/J=1.5\), the three values are \(0.029207\), \(0.041083\), and \(0.07150\)--\(0.07175\). At the first validated point above the history transition, the new history setting carries \(7.8\%\)--\(10.8\%\) of the optimized repetition budget; the newly active \(K^{(2)}\) settings carry \(11.3\%\)--\(19.6\%\). These are finite support changes, not optimizer-noise turn-ons.

No lower-shot history re-entry is found after the family crossover: immediately above \(H_3\) activation the history/instantaneous repetition ratio ranges from \(1.19\) to \(2.37\) across the tested fields, and at the upper \(K^{(2)}\) brackets it ranges from \(2.80\) to \(17.99\). Below \(\tau_{\rm family}\), history uses fewer repetitions and fewer nominal settings; above it, the instantaneous family is lower-shot while history remains lower-setting, giving a Pareto tradeoff.

The field dependence is approximately organized by \(\langle g\tau_{\rm family}\rangle=0.04268\) (CV \(3.07\%\)), \(\langle g\tau_{H_3}\rangle=0.06532\) (CV \(8.90\%\)), and \(\langle g\tau_{K^{(2)}}\rangle=0.10634\) (CV \(1.60\%\)). These are model- and task-specific constitutive regularities, not universal constants. The outer lag/support/shared-shot search remains a screened, refined, continued, and cross-seeded engineering optimization.

\suppnote{Internal structure of Ising history and independent spectral audit}
\label{supp:note-ising-internal}

\suppsubsection{Two distinct history architectures}

At the nominal point \(g/J=1.5\), the two-lag history landscape contains a local basin near \((Js_1,Js_2)=(0.06,0.20)\) and a mixed local/nonlocal basin near \((0.06,0.70)\). In the residual-first mechanism diagnostic their costs exchange order near
\begin{equation}
J\tau_{\rm arch}
\simeq
0.05840.
\label{supp:eq:ising-h3-architecture-crossing}
\end{equation}

where the directly evaluated repetition counts differ by only about \(6.6\times10^{-5}\) relatively. This internal architecture crossing is distinct from the final task-aware family, \(H_3\)-activation, and \(K^{(2)}\)-activation boundaries.

After removing the common present-time information, the first history layer at \(Js_1\simeq0.06\) has overlap \(0.969\) with the current subspace and mean principal angle about \(9.5^\circ\): in both branches it is mainly a local velocity coordinate. We then remove the direction already supplied by that first historical measurement. Table~\ref{supp:tab:ising-h3-geometry} shows what the second measurement adds.

\begin{table}[t]
\centering
\small
\begin{tabular}{lccc}
\hline\hline
\(H_3\) branch
&
\((Js_1,Js_2)\)
&
\(K^{(2)}\) overlap
&
mean angle
\\
\hline
Local / jet-like
&
\((0.06,0.20)\)
&
0.724
&
\(30.8^\circ\)
\\
Mixed local/nonlocal
&
\((0.06,0.70)\)
&
0.114
&
\(74.1^\circ\)
\\
\hline\hline
\end{tabular}
\caption{
Layer-resolved operator geometry of the two nominal \(H_3\) basins after
removing the common present-time information and conditioning the second
history layer on the first. A high overlap with \(K^{(2)}\) identifies a
more derivative-like second coordinate, while the long branch points mainly
outside the direct second-derivative subspace.
}
\label{supp:tab:ising-h3-geometry}
\end{table}

Thus the local second node remains substantially derivative-like, whereas the long second node points mainly outside the direct \(K^{(2)}\) space. At the architecture crossing the local branch also has the better intrinsic closure, approximately \(0.447\%\) versus \(0.682\%\), but the worse normalized Gram conditioning, \(2.90\times10^4\) versus \(1.57\times10^3\). The lower residual and lower metrological burden therefore favor different branches, illustrating why residual minimization alone is not the task-aware objective.

The lag search must consequently screen both local and nonlocal basins. A purely local search can miss the lower-cost long-memory branch; this is why the final outer search combines broad screening, local refinement, continuation, and cross-seeding.

\suppsubsection{Independent Solver-B audit}

Solver B reconstructs the spin operators independently and uses fourth-order Gauss--Magnus propagation with \(J\Delta t=5\times10^{-4}\). Ten scientifically decisive constraints around the nominal family, \(H_3\), and \(K^{(2)}\) gates are recomputed independently. Across those tests, the largest target-operator norm difference is \(1.133\times10^{-10}\), the largest record-operator difference is \(6.224\times10^{-11}\), and the spectral-distance upper values differ by at most \(2.651\times10^{-8}\). The largest cutting-plane upper-minus-lower gap is below \(9.754\times10^{-8}\), every Solver-A interval overlaps the corresponding Solver-B interval, and step halving changes the longest audited propagators by at most \(2.249\times10^{-14}\).

This audit validates the microscopic operator geometry beneath the final representation map. It does not turn the outer lag, support, coefficient, and shared-shot search into a theorem-level global optimization, and it is a numerical/model validation rather than an experiment.

\suppnote{Cross-platform geometry and complete numerical audit}
\label{supp:note-mechanism}

\suppsubsection{Common geometric diagnostics}

To compare history with direct derivatives, we first remove the operator directions already contained in the present-time span. If \(Q_A\) and \(Q_B\) are Hilbert--Schmidt orthonormal bases for two conditioned subspaces, their principal angles are defined by
\begin{equation}
\sigma_i
\left(
Q_A^\dagger Q_B
\right)
=
\cos\theta_i,
\label{supp:eq:principal-angle-definition}
\end{equation}

and we report the mean squared-cosine overlap and mean angle as
\begin{equation}
\mathcal O(A,B)
=
\frac{1}{q}
\sum_{i=1}^{q}
\cos^2\theta_i,
\quad
\overline{\theta}
=
\frac{1}{q}
\sum_{i=1}^{q}
\theta_i,
\label{supp:eq:subspace-overlap}
\end{equation}

where \(q\) is the number of compared principal directions.

Temporal redundancy is quantified by normalizing the conditioned history operators, \(\widehat B_a=B_a/\|B_a\|_{\rm HS}\), and forming their Gram matrix,
\begin{equation}
G_{ab}
=
\left\langle
\widehat B_a,
\widehat B_b
\right\rangle_{\rm HS}.
\label{supp:eq:history-gram}
\end{equation}

We report its condition number,
\begin{equation}
\kappa_{\rm Gram}
=
\frac{
\lambda_{\max}(G)
}{
\lambda_{\min}^{+}(G)
},
\label{supp:eq:gram-condition}
\end{equation}

Here \(\lambda_{\min}^{+}\) is the smallest retained positive eigenvalue of the Gram matrix. A high overlap and small angles mean that history largely measures the same directions as the retained local derivatives; a low overlap and large angles identify information missing from those particular derivative layers. A large \(\kappa_{\rm Gram}\) signals near-redundancy and possible statistical amplification. Geometry alone does not select the state: the constitutive margin and metrological burden must still be included.

The resulting overlaps and condition numbers are compared in Table~\ref{supp:tab:crossplatform-geometry}.

\begin{table}[t]
\centering
\small
\begin{tabular}{lccc}
\hline\hline
Case
&
\(\mathcal O_{\mathcal J}\)
&
\(\mathcal O_{K^{(2)}}\)
&
\(\kappa_{\rm Gram}\)
\\
\hline
BH local
&
0.9426
&
0.9148
&
\(3.920\times10^5\)
\\
BH practical
&
0.3133
&
0.2712
&
\(2.306\times10^3\)
\\
Ising \(J\tau=0.05\)
&
0.952
&
0.583
&
\(1.00\times10^4\)
\\
Ising \(J\tau=0.10\)
&
0.983
&
0.843
&
\(1.08\times10^5\)
\\
Ising \(J\tau=0.15\)
&
0.995
&
0.970
&
\(4.80\times10^6\)
\\
\hline\hline
\end{tabular}
\caption{
Common operator-geometric diagnostics for the two platforms. The
Bose--Hubbard ``local'' pair is a deliberately derivative-like reference,
while the ``practical'' pair is the history retained by the final
Bose--Hubbard architecture. The Ising rows come from the equal-coordinate
mechanism scan and are used to diagnose the evolution of the history
geometry, not as replacements for the final task-aware phase map.
}
\label{supp:tab:crossplatform-geometry}
\end{table}

For the Bose--Hubbard practical lags, the mean principal angles are \(56.03^\circ\) for the current layer and \(58.68^\circ\) for the conditioned second layer; the deliberately local pair gives only \(13.81^\circ\) and \(16.79^\circ\). The practical history is therefore far from the first two derivative layers and is about \(170\) times better conditioned than the local pair. Using the characteristic rates \(\Omega_1/J=23.70345\) and \(\Omega_2/J=21.39468\), the corresponding products are \(\Omega_1s_1\simeq2.37\) and \(\Omega_2s_2\simeq4.28\), confirming that these lags are not infinitesimal. In the Ising mechanism scan, the selected equal-coordinate lags move from approximately \((0.075,0.260)\) at \(J\tau=0.05\) to \((0.025,0.0525)\) at \(J\tau=0.15\), while the history approaches the derivative jet and its conditioning deteriorates by nearly three orders of magnitude.

The cross-platform conclusion is that no single geometric diagnostic selects the cheapest state: the operator directions, remaining constitutive margin, and amplification of measurement noise must be considered together. History is advantageous when it contributes useful finite-time directions at acceptable conditioning; direct dynamical observables become preferable when history mainly reconstructs the same local derivative information through poorly conditioned finite differences.

\suppsubsection{Validation and logical status}

For reference, the final numerical validation layers are collected in one place:

\begin{table}[t]
\centering
\small
\begin{tabular}{p{0.31\linewidth}p{0.61\linewidth}}
\hline\hline
Check & Result \\
\hline
BH continuous domain
&
Eight response tables; \(125\)-point sensitivity lattice plus interior
checks; \(20\%\) sensitivity inflation; intrinsic upper certificate
\(0.7304\%\)
\\
BH metrology
&
\(3.786\times10^7\) shared idealized repetitions for the full declared
continuous task
\\
BH blind validation
&
\(349{,}920\) predictions; maximum error \(0.1278\%\); worst-event
99th percentile \(0.1148\%\)
\\
BH Solver-B convergence
&
maximum unitarity defect \(1.517\times10^{-14}\); maximum step-halving
event difference \(2.002\times10^{-13}\)
\\
Ising phase map
&
\(\tau_{\rm family}<\tau_{H_3}<\tau_{K^{(2)}}\) at every tested
\(g/J\)
\\
Ising independent audit
&
10 selected transition constraints; maximum spectral-distance upper
difference \(2.651\times10^{-8}\); all Solver-A/Solver-B certified
intervals overlap
\\
Ising Solver-B convergence
&
maximum audited step-halving propagator difference
\(2.249\times10^{-14}\)
\\
Outer architecture search
&
screened, refined, continued, and cross-seeded numerical engineering
optimum; no theorem-level global-optimality claim
\\
\hline\hline
\end{tabular}
\caption{
Final numerical validation ledger. The rows deliberately separate
continuous-domain certification, holdout validation, numerical convergence,
and outer representation optimization because these tests establish
different kinds of reliability.
}
\label{supp:tab:final-validation-ledger}
\end{table}

The logical status of the results is deliberately separated. The operator-span determinacy theorem and exact global ambiguity relation are finite-dimensional analytic statements. The generic preparation-tube inequalities are rigorous sufficient certificates. The continuous Bose--Hubbard parameter-box statement is an engineering certificate under a validated numerical sensitivity envelope, not interval arithmetic. The Bose--Hubbard holdout and Ising Solver-B comparison are independent numerical/model validations, not experimental validations. The outer architecture searches are screened and cross-seeded engineering optima, and the Ising scaling numbers are model-specific constitutive regularities.

The resulting operational picture is simple. The detailed microscopic theory is used beforehand to select the record, calibrate the constitutive response, and establish its validity domain. Routine operation then uses only the selected measured state and the already calibrated response. The thermodynamic state is therefore not a universal smallest description of the microscopic system; it is the experimentally economical description sufficient for the declared task.

\end{document}